 \documentclass[final,3p,times,twocolumn]{elsarticle}

\usepackage{amssymb}
\usepackage{color}
\usepackage{ulem}

\journal{Journal of Computational Physics}

\def\vector#1{\mbox{\boldmath $#1$}}

\begin{document}

\begin{frontmatter}



\title{Validation of Radiative Transfer Computation with Monte Carlo Method for Ultra-Relativistic Background Flow}


\author[tohoku,tokyo]{Ayako Ishii}
\author[tohoku]{Naofumi Ohnishi}
\author[caltech]{Hiroki Nagakura}
\author[riken,ithes]{Hirotaka Ito}
\author[waseda]{Shoichi Yamada}

\address{}
\address[tohoku]{Department of Aerospace Engineering, Tohoku University, Sendai 980-8579, Japan}
\address[tokyo]{Research Center for the Early Universe, School of Science, University of Tokyo, Tokyo 113-0033, Japan}
\address[caltech]{TAPIR, Walter Burke Institute for Theoretical Physics, Mailcode 350-17, California Institute of Technology, CA 91125, USA}
\address[riken]{Astrophysical Big Bang Laboratory, RIKEN, Wako 351-0198, Japan}
\address[ithes]{Interdisciplinary Theoretical Science (iTHES) Research Group, RIKEN,Wako, 351-0198, Japan}
\address[waseda]{Advanced Research Institute for Science and Engineering, Waseda University, Tokyo 169-8555, Japan}

\begin{abstract}
We developed a three-dimensional radiative transfer code 
for an ultra-relativistic background flow-field by using the Monte Carlo (MC) method 
in the context of gamma-ray burst (GRB) emission.
For obtaining reliable simulation results in the coupled computation of MC radiation transport with relativistic hydrodynamics 
which can reproduce GRB emission, 
we validated radiative transfer computation in the ultra-relativistic regime 
and assessed the appropriate simulation conditions.
The radiative transfer code was validated through 
two test calculations: 
(1) computing in different inertial frames and 
(2) computing in flow-fields with discontinuous and smeared shock fronts. 
The simulation results of the angular distribution and spectrum were compared 
among three different inertial frames and in good agreement with each other. 
If the time duration for updating the flow-field was sufficiently small 
to resolve a mean free path of a photon into ten steps, 
the results were thoroughly converged.
The spectrum computed in the flow-field with a discontinuous shock front 
obeyed a power-law in frequency whose index was positive 
in the range from 1 to 10~MeV.
The number of photons in the high-energy side decreased 
with the smeared shock front 
because 
the photons were less scattered immediately behind the shock wave due to the small electron number density.
The large optical depth near the shock front was needed for obtaining high-energy photons through bulk Compton scattering. 
Even one-dimensional structure of the shock wave could affect 
the results of radiation transport computation.
Although we examined the effect of the shock structure on the emitted spectrum with a large number of cells, 
it is hard to employ so many computational cells per dimension in multi-dimensional simulations.
Therefore, a further investigation with a smaller number of cells is required 
for obtaining realistic high-energy photons with multi-dimensional computations.

\end{abstract}

\begin{keyword}
Gamma-ray burst, Relativistic jet, Radiative transfer, Monte Carlo method
  


\end{keyword}

\end{frontmatter}


\section{Introduction}
\label{}
Relativistic radiation hydrodynamics computation is widely used 
in the field of high-energy astrophysics, 
e.g., gamma-ray bursts (GRBs), supernovae, accretion discs, and active galactic nuclei.
It is difficult to analytically solve the radiative transfer equation since it is a seven-dimensional equation; 
therefore, some approximated methods have been proposed to obtain physically reasonable solutions 
\cite{pomraning, webb, kirk}.

The moment method, which wraps up angular dimensions of photons, 
is frequently used in radiation hydrodynamics computation \cite{pomraning}. 
However, it is valid only for a closely isotropic radiation field 
(that is, an optically thick regime), 
unless an appropriate closure is given.
For a highly relativistic regime, 
it is inadequate because the radiation field holds strong anisotropy as a result of the beaming effect.
The moment method was also formalized in the relativistic regime, 
but it is applicable for simple limiting cases \cite{anderson}.
The discrete-ordinate method, which solves the radiative transfer equation with the finite differencing of 
direction components, is suitable for an anisotropic regime \cite{tominaga}. 
However, it generally requires huge computational costs for resolving six-dimensional phase space. 
The Monte Carlo (MC) method is useful to statistically obtain a solution of a multi-dimensional integro-differential equation 
and then to solve the radiative transfer equation including scattering process. 
It is appropriate in an optically thin and a mildly scattering regime 
such as a radiation field accompanying a relativistically expanding jet, resulting in GRB emission.

GRBs are highly energetic explosion phenomena 
in which extremely large amounts of energy are emitted in a few seconds to minutes. 
In particular, long GRBs of duration greater than $\sim$2 s 
are thought to occur in association with 
relativistic jets formed around collapsing massive stars and have a collimated configuration, 
while short GRBs of duration less than $\sim$2 s 
are considered to result from a binary merger of compact objects, 
e.g., neutron star--neutron star or neutron star--black hole. 
The prompt gamma-ray emission produced in internal shocks of an ultra-relativistic jet is typically interpreted 
as a result of synchrotron radiation from shocked-accelerated electrons \cite{rees, sari}. 
However, the internal shock model could provide insufficient radiation efficiency for explaining GRB emission 
\cite{kobayashi,mimica1, mimica2}. 
On the other hand, the effect of a photosphere position at which thermal photons emerge has been discussed 
for the thermal component of GRB spectra \cite{eichler, meszaros}.
Such a photospheric emission model, in which thermal photons are Comptonized at the shock front, 
has high radiation efficiency for GRBs.
The structure of relativistic jets, which develops with time, has also been studied 
through multi-dimensional relativistic hydrodynamics simulations 
in the context of GRBs \cite{aloy1, aloy2, zhang, mizuta1, nagakura, matsumoto}, 
and light curves and spectra have been estimated based on such simulations 
\cite{nagakura, lazzati, mizuta2, martinez1, martinez2}. 
Although the observed spectra can be characterized by a broken power-law shape \cite{briggs}, 
their spectral properties have not been accurately reproduced in the numerical works. 

The MC radiative transfer computations have been implemented 
in the relativistic regime relevant to GRBs \cite{lucy, beloborodov, ayako1, shibata, ayako2}. 
Numerical studies with the MC technique have been also reported 
for explaining radiation and neutrino transport 
in supernova explosions 
\cite{janka0, janka1, janka2, keil, abdikamalov, maeda, suzuki}. 
Some observations indicate the GRB spectra includes a thermal component \cite{ryde, thone}; 
therefore, the radiation transport of thermal photons produced at the photosphere has been paid attention, 
and the non-thermal feature of the spectra was obtained 
by overlapping thermal spectra 
with various escaped angles at different time \cite{peer}. 
The non-thermal spectra were also obtained by taking into account the gradual energy dissipation 
by magnetic reconnection \cite{giannios1, giannios2}. 
On the other hand, the structure of an ultra-relativistic shock wave in the self-similarly expanding fluid 
has been numerically investigated \cite{nakayama}, 
and the high-energy component of GRB spectra could be explained by bulk Compton scattering in such a shock wave 
by using MC computations \cite{takagi, ohtani}.  
The bulk Compton scattering occurs when photons traveling 
across the shock wave collide with relativistic electrons. 
The non-thermal spectra of GRBs have been also explained through bulk Comptonization in an ultra-relativistic jet 
with some shells of various flow velocities \cite{ito, ito2}. 

Past works with multi-dimensional relativistic hydrodynamics simulations 
showed that structure of a relativistic jet exhibits highly inhomogeneous developing, 
which can affect the observed spectra 
\cite{aloy1, aloy2, mizuta1, nagakura}. 
Although some radiative transfer simulations were conducted with an ultra-relativistic steady flow-field, 
those with a time-dependent flow-field may have a significant impact on 
the detailed analysis of GRB spectra.
Relativistic radiation hydrodynamics simulations were implemented with the MC radiative transfer; 
however, 
some of them 
do not appropriately take into account the feedback from interaction of radiation with 
flow-field matter \cite{noebauer, ito3}, 
while 
others 
do not sufficiently perform the test calculations for the ultra-relativistic flow-field \cite{roth}. 
On the other hand, radiation hydrodynamics calculations for predicting the emission of internal magnetized shocks 
including the feedback 
of the radiation on the dynamics were 
examined in the relativistic regime without the MC technique \cite{mimica3}.
The coupling of the MC radiative transfer 
with relativistic hydrodynamics has 
not 
been sufficiently performed 
because of computational difficulties as introduced in Sec.~\ref{sec:difficulty}. 
In this paper, we 
show the validation of the formalism to include the MC method 
in the ultra-relativistic flow-field 
and examine the appropriate simulation conditions such as time interval for developing the flow-field 
and spatial resolution for obtaining converged results. 

The remainder of this paper is organized as follows.
We introduce the difficulty in performing MC radiative transfer computation 
in the ultra-relativistic flow-field in Sec.~\ref{sec:difficulty}, 
and present the numerical method in Sec.~\ref{sec:numerical}. 
The validation of the radiative transfer computation in different inertial frames is presented 
in Sec.~\ref{sec:dif_frame}, 
and the effect of the flow-field resolution on radiation transport is discussed in Sec.~\ref{sec:smeared_discontinuous}.
We summarize this paper in Sec.~\ref{sec:conclusion}.

\section{Difficulties in performing Monte Carlo radiative transfer computation coupled with ultra-relativistic hydrodynamics}
\label{sec:difficulty}
Some MC methods for radiation transport have been developed 
for the coupling with non-relativistic hydrodynamics computation in previous works as follows.
Implicit MC (IMC) schemes 
have 
been employed for efficient computations in 
optically thick systems, 
and 
they allow employing larger time steps, 
$\Delta t$, 
in the numerical simulation of the flow-field 
\cite{fleck, mcclarren}.
The diffusion approximation which can reduce a computational cost is reasonable in opaque regions, 
so a hybrid method consisting of IMC and MC diffusion method has been developed \cite{gentile, wollaeger,cleveland}.
Since the MC method includes statistical errors with a small number of samples, 
some techniques for reduction of statistical noise 
have 
been investigated 
with a moderate computational cost \cite{densmore1, densmore2}.
These techniques are actually effective in the non-relativistic regime. 

\begin{figure}[h]
\begin{center}
  \includegraphics[width=0.8\linewidth]{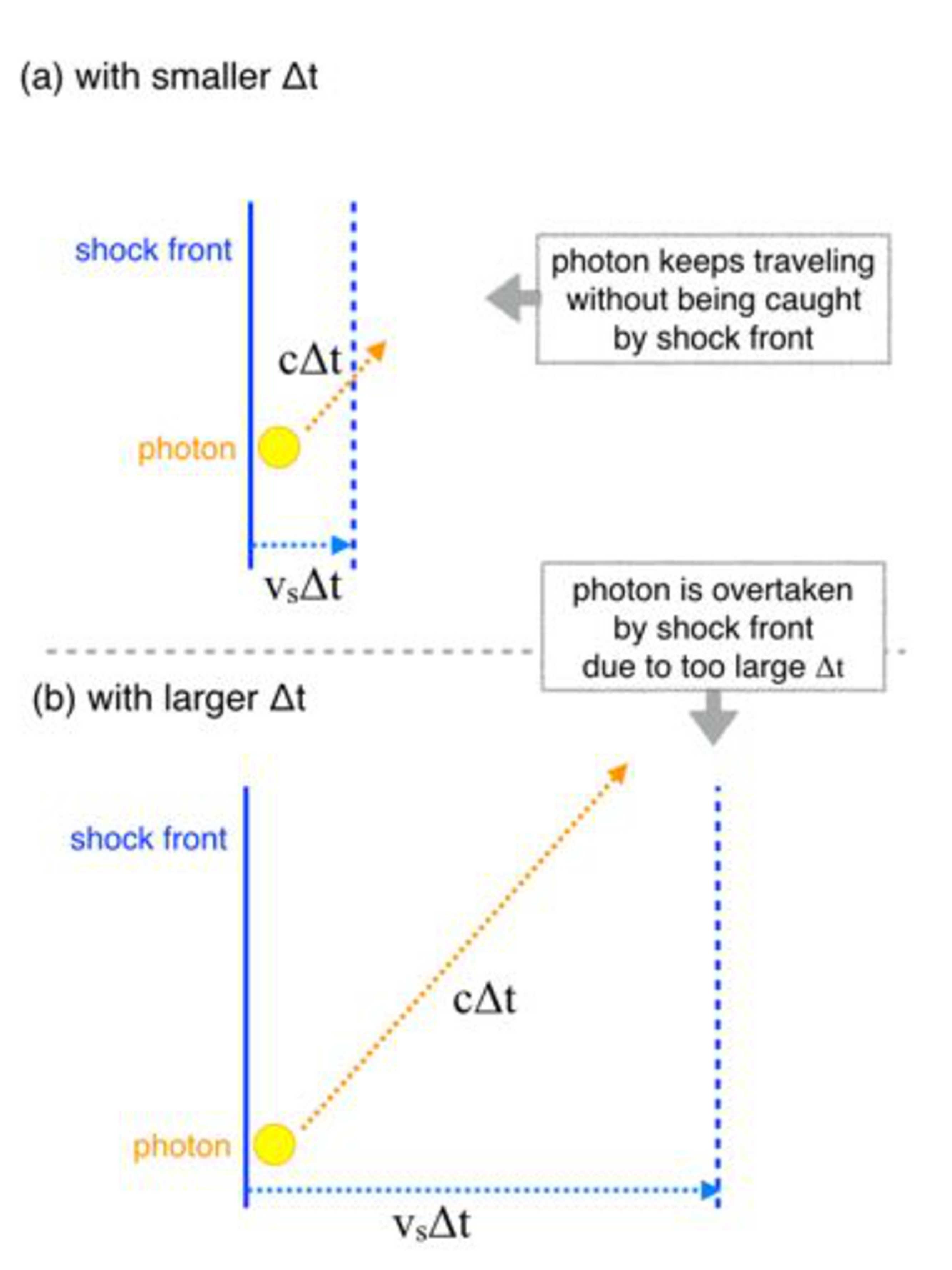}
  \caption{Problem of photon transfer in an ultra-relativistic flow-field. 
(a) A photon keeps traveling without being caught by the shock front in the situation with smaller $\Delta t$, 
and (b) a photon is overtaken by shock front because of an excessively large $\Delta t$.}
  \label{fig:rela_problem}
\end{center}
\end{figure}

In the ultra-relativistic regime 
(Lorentz factor $\Gamma \gtrsim 100$), 
however, 
since the velocity of matter is almost the same as the speed of light, 
computation with an excessively large $\Delta t$ for developing the flow-field leads to 
false judgment on whether a photon crosses the shock front (Fig.~\ref{fig:rela_problem}).
Photons in the upstream side of the shock wave are scattered with a different probability 
from those in the downstream side.
If the Compton scattering is considered, scattered photons not only shift their directional angles 
but also undergo energy exchange with matter. 
Therefore, the simulation of radiation transport could not produce accurate results in 
directional-angle distribution and an energy spectrum 
with a mistake in judgment on whether a photon crosses the shock front due to the large $\Delta t$. 
A small $\Delta t$ is essentially necessary in the ultra-relativistic regime, 
so the IMC scheme for a large $\Delta t$ might be meaningless, 
and it is controvertible to develop the IMC scheme for highly relativistic situations.
Since a strongly anisotropic flow should be taken into account in the ultra-relativistic regime because of the beaming effect, 
the diffusion approximation cannot be applicable.

In computations including feedback from interaction between photons and flow-field matter, 
a huge computational cost is needed.
Since the flow-field profile is affected by the radiation transport, 
the post-process computation of the radiation transport 
in the steady background flow-field computed in advance is not adequate. 
Therefore, the computations of radiative transfer and hydrodynamics should be alternately performed.
Furthermore, a number of photons should be put in each cell of the hydrodynamics computation 
for obtaining converged simulation results, 
however, it leads to a huge-computational-cost problem. 
Thus, some techniques for reduction of statistical noise with a small number of samples as stated above are required, 
and the coupled computation may be feasible if 
restricted to some regions in the flow-field.

Moreover, numerical diffusion is inevitable in hydrodynamics simulation, 
which especially affects shock structure. 
Photons near the shock front are scattered with the bulk-Compton-scattering process 
and obtain energy from relativistic electrons in the flow-field 
according to the local flow velocity and density.
The local flow velocity affects the obtained energy from the scattering, 
and density is related to the probability of the scattering.
The shock structure, therefore, affects the spectrum resulting from the radiative transfer computation. 
Unphysical finite width of a smeared shock front due to numerical diffusion depends on spatial resolution of the hydrodynamics computation. 
Appropriate simulation conditions should be 
assessed 
such as the time interval for updating the background flow-field 
and the spatial resolution 
for feasible computations of radiation transport 
before the coupled computation of the MC radiative transfer with ultra-relativistic hydrodynamics 
can be performed.

In the present study, we construct radiative transfer algorithm 
for an ultra-relativistic background flow-field 
with consistent transformation between a comoving frame (CMF) and an observer frame (OBF), 
and 
assess the interval of time-step values which yield accurate computations. 
The effect of the spatial resolution of the background flow-field on the radiative transfer computation 
is 
also discussed. 

\section{Numerical method}
\label{sec:numerical}
We developed a simulation code of radiation transport for a highly relativistic background 
with the MC technique.
The radiative transfer equation in a certain inertial frame that takes scattering into account is expressed as follows \cite{pomraning}:
\begin{eqnarray}
  \label{eq:rad_transfer}
  \hspace{-10mm}
  \lefteqn{ \left( \frac{1}{c} \frac{\partial}{\partial t} + \vector{\Omega} \cdot \nabla \right)
    I \left( \vector{r}, \vector{\Omega}, \nu, t \right) = j \left(\vector{r}, \vector{\Omega}, \nu, T \right)} \\ \nonumber
  \hspace{-10mm}
  \lefteqn{ - k \left(\vector{r}, \vector{\Omega}, \nu \right) 
  \rho \left(\vector{r},t \right) I \left(\vector{r},\vector{\Omega},\nu,t \right)} \\ \nonumber
  \hspace{-10mm}
  \lefteqn{ + \rho \left( \vector{r}, t \right) \int^{\infty}_0  \int_{4 \pi} 
  [ (\nu / \nu') \sigma \left(\vector{r}, \vector{\Omega}' \cdot \vector{\Omega}, \nu' \rightarrow \nu \right) 
  I \left( \vector{r}, \vector{\Omega}', \nu', t \right)} \\ \nonumber
  \hspace{-10mm}
  \lefteqn{ - \sigma \left(\vector{r}, \vector{\Omega} \cdot \vector{\Omega}', \nu \rightarrow \nu' \right)
  I \left( \vector{r}, \vector{\Omega}, \nu, t \right) ] d \vector{\Omega}' d \nu',}
\end{eqnarray}
where $I( \vector{r},\vector{\Omega}, \nu, t)$ is the specific intensity, 
which is a function of the position vector $\vector{r}$, 
the traveling direction vector $\vector{\Omega}$, photon frequency $\nu$, and time $t$. 
The symbols $c$ and $\rho$ denote the speed of light and density, respectively. 
The absorption and scattering cross-sections are denoted by $k$ and $\sigma$, respectively. 
The emissivity $j$ depends on the matter temperature $T$.
In the integrand, $\vector{\Omega}$, $\vector{\Omega}'$, $\nu$, and $\nu'$ denote 
the scattered direction, incident direction, scattered frequency, and incident frequency, respectively. 
The fluid velocity varies depending on the position; 
therefore, the cross-sections are dependent on $\vector{r}$ in an arbitrary inertial frame.

Photons are thermally emitted in the flow-field 
and travel in a medium in which scattering opacity is assumed to be greater than absorption opacity. 
Thomson and Compton scattering processes are taken into account. 
We adopt the MC technique to solve the radiative transfer equation as in the previous works \cite{ayako1,ayako2}.

By tracking a large number of particles, 
we can obtain an approximate solution of the equation with reasonable computational costs. 
A cluster of photons (called as `packet') having single frequency is considered as a sample particle. 
The packet has energy of $\epsilon (\nu) = n h_p \nu$, 
where $h_p$ and $n$ are Planck constant and the number of photons in the packet, respectively.
Emission power $\dot{e}$ at any position is described as 
the spectral integration of the emission coefficient in the CMF, $j_0$, given by 
\begin{equation}
\dot{e} = \int^{\infty}_{0} j_0 (\nu_0, T) d \nu_0, 
\end{equation}
where $\nu_0$ is the photon frequency in the CMF; 
the initial value of $\nu_0$ is set randomly 
as weighted for Planck energy distribution in the CMF. 
The holding energy for single packet in the CMF, $\epsilon_0$, 
can be obtained as product of $\dot{e}$, the total cell volume, and a certain time duration 
divided by the total number of emitted packets during the time interval, $N$.
Since $\epsilon_0$ is the energy in the CMF, 
we need to Lorentz-transform it to that in the OBF. 
The frequency in the CMF, $\nu_0$, is transformed to that in the OBF as well.
The initial direction of the photon is also set by two random numbers 
with the assumption of isotropic emission in the CMF; 
the initial direction vector 
$\vector{\Omega}_0 = ({\rm sin}\theta {\rm cos}\phi, {\rm sin}\theta {\rm sin}\phi, {\rm cos}\theta)$ 
can be expressed by 
\begin{eqnarray}
{\rm cos}\theta &=& 1 - 2 R_1,\\
\phi &=& 2 \pi R_2,
\end{eqnarray}
where $R_1$ and $R_2$ are random numbers for the direction. 
The traveling direction in the CMF, $\vector{\Omega}_0$, is transformed 
to obtain the one in the OBF, $\vector{\Omega}$, as follows: 
\begin{equation}
\vector{\Omega} = \frac{ \left[ \vector{\Omega}_0 + \vector{v}/c \left( \Gamma_f + \Gamma_f^2/(\Gamma_f + 1) 
\vector{\Omega}_0 \cdot  \vector{v}/c \right) \right] }{\Gamma_f (1 + \vector{\Omega}_0 \cdot \vector{v} /c )},
\end{equation}
where $\Gamma_f$ is the Lorentz factor of the flow velocity.
Using the direction $\vector{\Omega}$ and the flow velocity $\vector{v}$, 
the frequency $\nu$ and packet energy $\epsilon$ in the OBF are given by 
\begin{eqnarray}
\label{eq:nu_trans}
\nu = \nu_0 \frac{\sqrt{1-(v/c)^2}}{1-\vector{\Omega} \cdot \vector{v}/c},\\
\epsilon = \epsilon_0 \frac{\sqrt{1-(v/c)^2}}{1-\vector{\Omega} \cdot \vector{v}/c}.
\end{eqnarray}

Photons are categorized into two energy groups because they have different cross-sections 
depending on their energy regime. 
The lower-energy photons are called as `optical ray', 
and the higher-energy ones are expressed by `gamma ray'. 
The threshold of two groups is set to 10 keV.
The scattering cross-section for optical ray is Thomson scattering one, $\sigma_T$. 
The electron number density, $n_e$, is estimated by assuming that 
the species involved in the background is only fully-ionized helium gas in this paper, 
so the mass scattering cross-section in the CMF, $\sigma_0$, for optical ray can be calculated 
as $\sigma_0 = \sigma_T n_e / \rho$. 
In the Thomson scattering, the energy of the incident photon in the CMF is fully transferred to the scattered one. 
The scattered angle and direction are determined 
by assuming unpolarized light and using a phase function $(3/4) ({\rm cos}^2 \Theta + 1)$, 
where $\Theta$ is the scattering angle.

Gamma-ray photons are scattered in the Compton regime.
Thus, the corresponding Compton cross-section is employed for them.
The energy of the incident gamma ray is transferred to the scattered light and matter, 
and the photon loses its own energy to become an optical ray 
through multiple scatterings. 
The scattering cross-section in the CMF can be estimated by integrating the Klein-Nishina (K--N) formula.
The scattering angle, $\Theta$, can be determined by a random number, $R_3$, employing the K--N formula; 

\begin{equation}
R_3 = \frac{1}{\sigma_{KN}(E_0)} \int^{\Theta}_0 \frac{\partial \sigma_{KN} (E_0, \Theta')}{\partial \Theta'} d \Theta',
\end{equation}
where $E_0$ and $\sigma_{KN}$ are the energy of the incident photon 
in the CMF 
and K--N scattering cross-section, respectively. 
Here, the differential K--N cross-section is expressed as follows: 
\begin{eqnarray}
\label{eq:K-N_eq}
\hspace{-20mm}
%
%
\frac{\partial \sigma_{KN}(E_0, \vector{\Omega})}{\partial \vector{\Omega}} &=& r_0^2 
\frac{1+ {\rm cos}^2\ \Theta}{2} \frac{1}{ [1+x(1- {\rm cos}\ \Theta )]^2}\\ \nonumber
& \times & \left( 1 + \frac{x^2 (1- {\rm cos}\ \Theta)^2}
{(1+ {\rm cos}^2\ \Theta) [1+ x (1- {\rm cos}\ \Theta) ] } \right),
\end{eqnarray}
where $x = E_0 / m_e c^2$ and $r_0 = e^2 / m_e c^2$. 
Here, $m_e$ and $e$ are the electron rest mass and the elementary charge, respectively.
The total K--N cross-section is obtained by integrating Eq.~(\ref{eq:K-N_eq}) over the solid angle \cite{heitler}. 
The angle $\Phi$ in a plane perpendicular to $\vector{\Omega}_0$ is randomly chosen 
by $\Phi = 2 \pi R_4$, 
where $R_4$ is a random number.
The energy of the incident packet is divided 
into 
the scattered light and the scattered electron, 
and the fraction of the energy of the scattered light is denoted as $f_c = 1 / \left[1 + E_0 / m_e c^2 (1-{\rm cos}\ \Theta) \right]$. 
So, the energy of the scattered light in the CMF, $E'_0$, is $E'_0 = f_c E_0$. 
Here, electron energy spectrum is not taken into account for simplicity. 

A free path of a photon in the CMF, $\delta s_0$, between the collisional events depends on the cross-section in the CMF. 
The optical depth of the matter corresponding to $\delta s_0$ is given by 
\begin{equation}
\tau_{fp} =  \sigma_0 \rho \delta s_0.
\end{equation}
The free path of the photon is determined by probabilistic manner using $\tau_{fp}$. 
The probability that the photon freely travels can be represented by
\begin{equation}
p(\tau_{fp}) = e^{-\tau_{fp}}.
\end{equation}
The uniform random number for the free path, $R_5$, equals to the integrated probability in the following form: 
\begin{equation}
R_5 = \int_0^{\tau_{fp}} p(\tau_{fp}') d \tau_{fp}',
\end{equation}
so that the optical depth is readily expressed with it;
\begin{equation}
\tau_{fp} = - {\rm ln} (1 - R_5).
\end{equation}
Therefore, the flying distance of the photon, $\delta s_0$, is given by 
\begin{equation}
\delta s_0 = - {\rm ln} (1 - R_5) / \sigma_0 \rho.
\end{equation}
Since photon transport is treated in the OBF, 
the free path in the CMF, $\delta s_0$, should be transformed to that in the OBF, $\delta s$, by 
\begin{equation}
\label{eq:trans_free-path}
\delta s = \delta s_0 \frac{\sqrt{1-(v/c)^2}}{1-\vector{\Omega} \cdot \vector{v}/c}.
\end{equation}
The time it takes for the packet to travel a distance $\delta s$ at light speed is $\delta t_p = \delta s / c$.
The time $t_p$ for the packet should be incremented to $t_p + \delta t_p$ between the events.

We validate our simulation code of radiative transfer 
through two test problems: 
(1) comparing the simulation results calculated in different inertial frames, and 
(2) evaluating the effect of the smeared shock front mimicking numerical diffusion on the radiative transfer simulation 
as a preliminary result for the coupled computation of radiative transfer with relativistic hydrodynamics.

\section{Comparison among different inertial frames}
\label{sec:dif_frame}
We performed radiative transfer simulations in a shock rest frame and two shock moving frames.
In the MC radiation transport computation, 
the part of collision with electrons is treated in the CMF, 
and that of transport of each photon is treated in the OBF.
Thus, the Lorentz transform between the CMF and OBF is performed. 
By comparing the results computed in the different frames, 
we verify that our simulation code can properly handle Lorentz transformation 
and determine the appropriate computational conditions 
for obtaining 
converged 
solutions.

\subsection{Simulation condition}
\begin{figure}[t]
\begin{center}
  \includegraphics[width=0.8\linewidth]{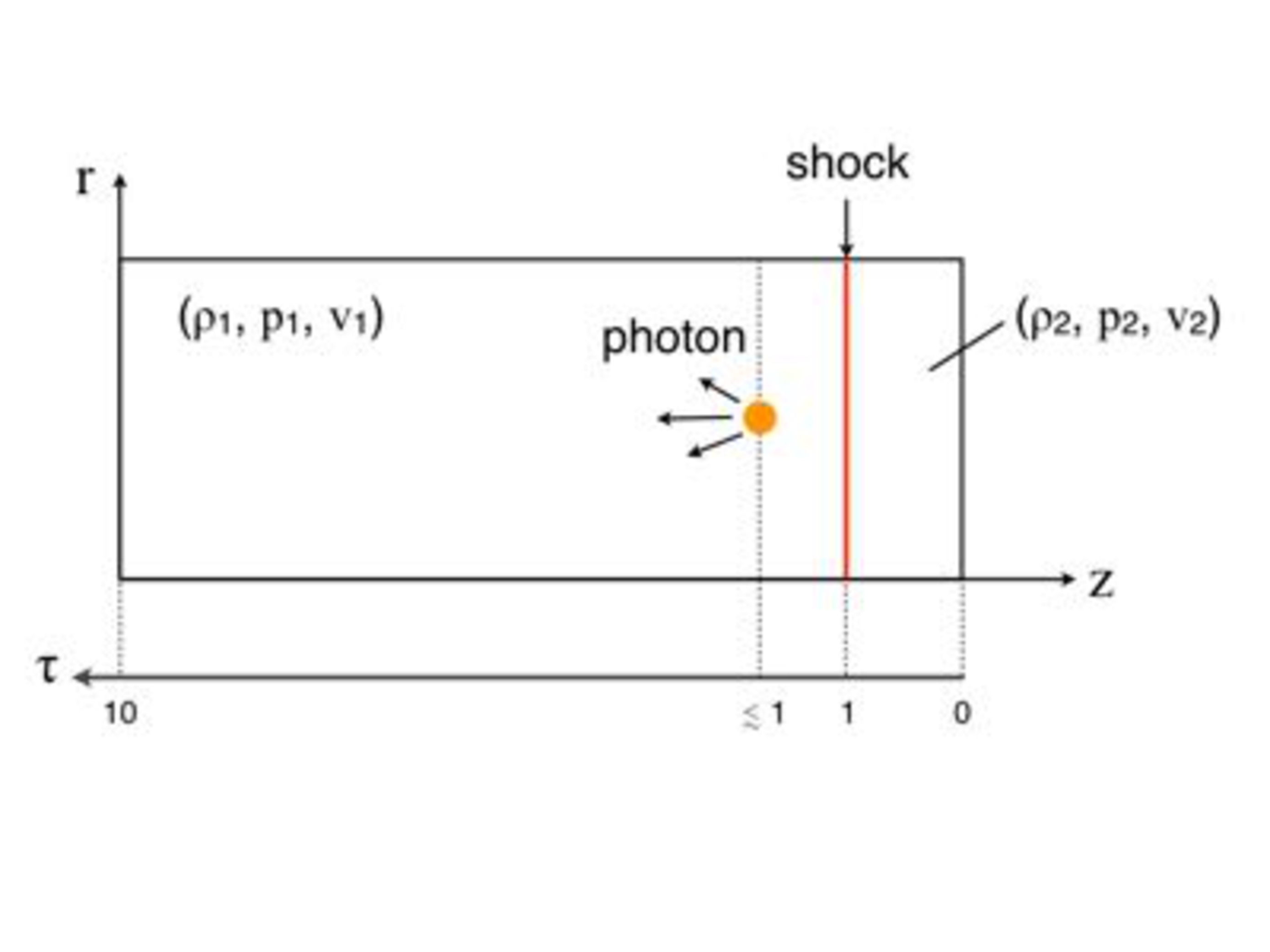}
  \vspace{-10 mm}
  \caption{Setting of a one-dimensional shock wave in the shock rest frame.}
  \label{fig:setting}
\end{center}
\end{figure}

A relativistic flow-field with a steady shock wave was set up, and 
a cylindrical coordinate system with one cell in the $r$-direction 
and two cells in the $z$-direction was adopted. 
The density $\rho_2$, pressure $p_2$, and flow velocity $v_2$ are the physical parameters 
at the upstream side of the shock, as shown in Fig. \ref{fig:setting}. 
Similarly, the physical parameters $\rho_1$, $p_1$, and $v_1$ are the values at the downstream side. 
These quantities satisfy the Rankine--Hugoniot (R-H) relations 
between the upstream and downstream sides of the shock wave; 
therefore, the cell interface in the $z$-direction represents the shock front. 
The width of the cell in the $z$-direction, $\Delta z_c$, is determined by the optical depth $\tau$ 
measured from the boundary at the upstream side along the $z$-axis as
\begin{equation}
\label{eq:tau_deltaz}
\tau = \rho \sigma_0 \Gamma_f \left(1- \frac{v_f}{c} \right) \Delta z_c,
\end{equation}
such that $\tau=1$ and $\tau=10$ at the upstream and downstream sides, respectively, as shown Fig.~\ref{fig:setting}. 
Here, the density $\rho$ and the scattering cross-section $\sigma_0$ are the values in the CMF, 
and $\Gamma_f$ is the Lorentz factor of the flow velocity $v_f$ in the shock rest frame. 
The widths of the computational cells in the shock moving frames can be fixed through the Lorentz contraction. 
To make the computational cells equivalent among different inertial frames, 
the length is measured by means of $\tau$. 
The width of the computational cell in the $r$-direction is set to 
be large enough not to affect simulation results.
All photons are emitted once at a pre-established initial time and 
at one point located at an optical depth $\tau \sim 1$ in the $z$-direction.
The location and timing of emitted photons are coincident on a space-time diagram among each frame. 
A lot of photons are tracked until the photons reach the boundary of the computational domain. 
We record directional angles and energy of the photons at the boundary 
and sum them for all the photons to obtain directional angular distribution and a spectrum.

The relativistic shock wave in the background can be described by solving the following relativistic R-H relations: 
\begin{eqnarray}
\left[\rho u^{z} \right] &=& 0, \\
\left[\rho h \left( u^{z} \right)^2 + p \right] &=& 0, \\
\left[\rho h u^{0} u^{z} \right] &=& 0,
\end{eqnarray}
where $u$, $h$, and $p$ are the four-velocity, specific enthalpy, and pressure, respectively. 
Here, $\vector{u} = \Gamma \left(1, v^1, v^2, v^3 \right)$, where $\Gamma$ is the Lorentz factor and $v^i$ is 
the three-velocity of the $i$th-direction component. 
The specific enthalpy is defined by $h = 1+p/ \rho + \epsilon$, where $\epsilon$ is the specific internal energy 
given by the equation of state for an ideal gas, 
$p = \left( \gamma - 1 \right) \rho \epsilon$.
A constant specific heat ratio $\gamma = 5/3$ is assumed.
The upstream quantities $\rho_2$, $p_2$, and $v_2$ and the shock speed $v_s$ are given; 
then, the downstream quantities $\rho_1$, $p_1$, and $v_1$ can be obtained by solving these relations. 
The physical quantities in the shock rest frame are adopted as summarized in Table~\ref{tab:butsuriryou}.
These quantities are employed by reference to the previous work 
on the relativistic hydrodynamics computation of a jet \cite{nagakura}, 
in which the flow-field transits from the optically thick to the optically thin regime. 
Here, the value of the flow velocity $v_2$ corresponds to $\Gamma_2 = 100$, 
where $\Gamma_2$ is the Lorentz factor of the upstream flow velocity in the shock rest frame, 
and $\Gamma_2$ is equal to the Lorentz factor of the shock speed $v_s$, $\Gamma_s$, 
in the rest frame of the upstream flow velocity.

The temperature $T$ of the matter is calculated 
from the equation of blackbody radiation by assuming the radiation equilibrium gas 
\begin{equation}
3 p_{\nu} = \frac{4 \sigma_s}{c} T^4, 
\end{equation}
where $\sigma_s$ is the Stefan--Boltzmann constant.
The emission coefficient and initial frequency of the packet are determined according to $T$, 
and electrons are assumed to be at rest in the CMF.
Here, $T$ in the downstream side of the shock wave corresponds to $1 \times 10^7\ {\rm K}$. 
The radiation pressure $p_{\nu}$ is assumed to be equal to the total pressure for simplicity. 

\begin{table}[t]
\begin{center}
 \caption{Physical quantities across the relativistic shock wave in the shock rest frame, the shock speed of which corresponds to $\Gamma_s = 100$.}
\label{tab:butsuriryou}
 \begin{tabular}{c|ccc}
 \hline
       & $\rho$ $[\rm{g/cm^3}]$ & $p$ $[\rm{dyn/cm^2}]$ & $v$ $[\rm{cm/s}]$ \\
 \hline
 upstream & $1.00 \times 10^{-11}$ & $1.00 \times 10^7$ & $-0.99994\ c$ \\
 \hline
 downstream & $1.16 \times 10^{-9}$ & $3.13 \times 10^{13}$  & $-0.65232\ c$ \\
 \hline
 \end{tabular} 
\end{center}
\end{table}

\subsection{Transformation of inertial frames}
\label{sec:trans_frame}
\begin{figure}[t]
\begin{center}
  \includegraphics[width=1.0\linewidth]{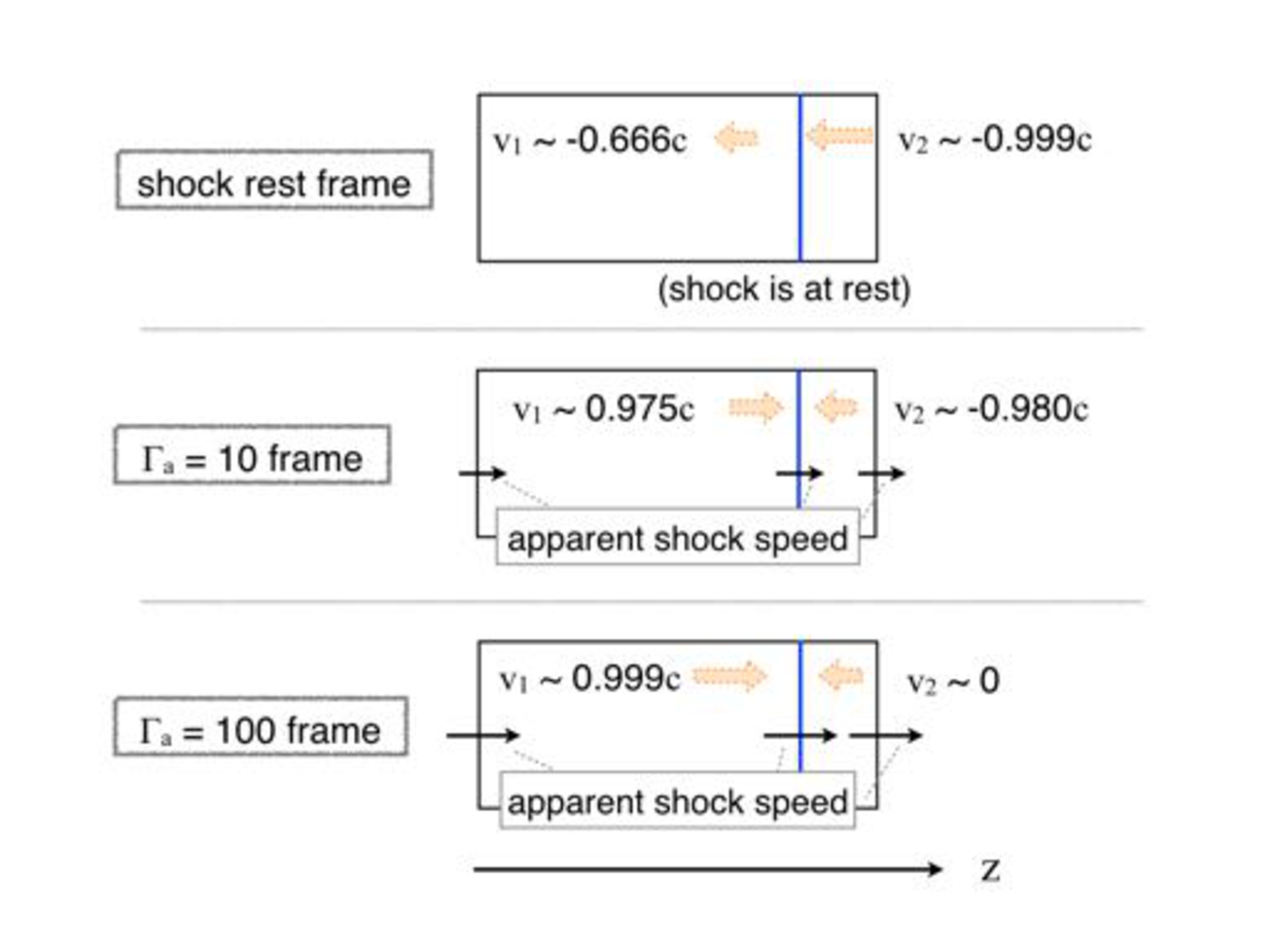}
  \vspace{-10 mm}
  \caption{Flow velocities and shock speed in each inertial frame. Thick and thin arrows denote flow velocities and shock speed, respectively.}
  \label{fig:sokudo}
\end{center}
\end{figure}

We transform the flow velocities from the shock rest frame to the shock moving frames 
in both the upstream and downstream sides of the shock wave 
to prepare different frames for the same shock-wave strength. 
The transformation is implemented according to the following equation \cite{schutz}:
\begin{equation}
\label{eq:velo_trans}
W' = \frac{W+v_a}{1+W v_a},
\end{equation}
where $W$ is the flow velocity in the shock rest frame, $W'$ is the flow velocity 
in the shock moving frame, 
and $v_a$ is the relative velocity between two frames (i.e., the apparent shock speed in the arbitrary frame). 
The flow velocities in each observer frame 
are shown in Fig.~\ref{fig:sokudo}.

We compare the simulation results for three different inertial frames in this paper. 
One is the shock rest frame, and the other two are shock moving frames 
corresponding to $\Gamma_a = 10$ and $100$, where $\Gamma_a$ is the Lorentz factor of  
the apparent shock speed, $v_a$. 
Here, the situation with $\Gamma_a = 100$ corresponds to the rest frame of the upstream flow velocity. 

To ensure that the computational grids on the space-time diagram 
are consistent between the shock rest frame and the shock moving frames, 
the computational grids move at the shock speed in the shock moving frames, 
and the shock front is located at the cell interface in each time step 
so that it has an exact discontinuity at all times. 
The shock front is assumed to be static during a time step $\Delta t$. 
The radiation transport is computed on the flow-field with the static shock front during the current $\Delta t$. 
Then, the position of the shock front is updated by the next $\Delta t$, 
and the radiation transport is computed on the updated flow-field with the static shock front again.
The shock front movement and the radiation transport are repeated alternately.

\subsection{Constraint of time interval}
\label{sec:const_dt}
\begin{figure}[t]
\begin{center}
  \includegraphics[width=0.8\linewidth]{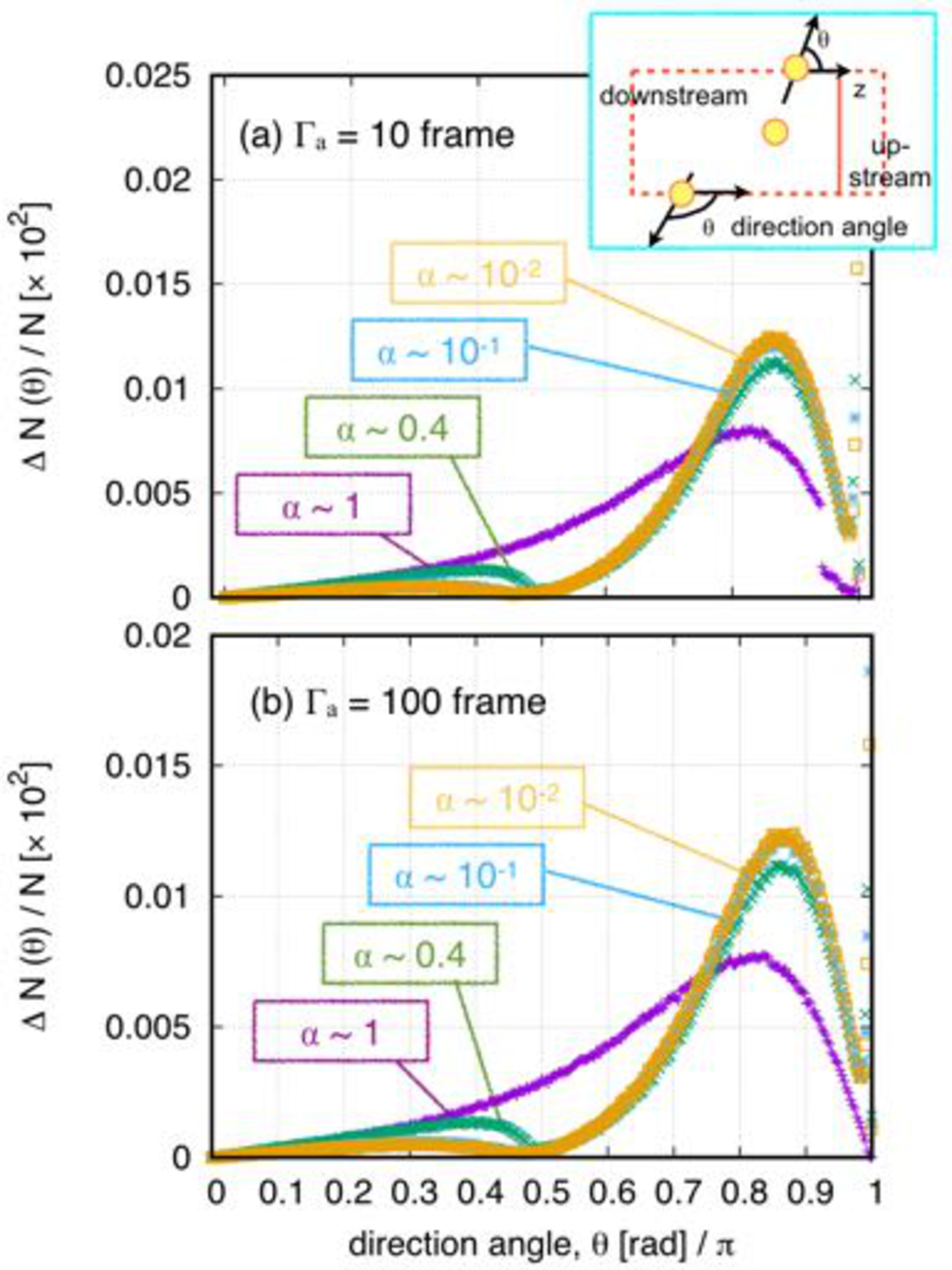}
  \caption{Angular distributions after transforming to the shock rest frame with various $\Delta t$ values. (a) $\Gamma_a = 10$ frame and (b) $\Gamma_a = 100$ frame.}
  \label{fig:direc_dt_010-100}
\end{center}
\end{figure}

We consider the required constraint of the computational time interval updating the flow-field, $\Delta t$, 
for 
obtaining converged solutions 
from the angular distribution of the photons escaping from the computational domain.
The angular distribution between the photon traveling direction and the $z$-axis 
with various $\Delta t$ values, which are calculated in the $\Gamma_a = 10$ frame 
with 
$10^6$ sample particles and transformed to the shock rest frame, are shown in Fig.~\ref{fig:direc_dt_010-100}~(a). 
For simplicity, we ignored Compton scattering and only took into account Thomson scattering 
with the low-frequency approximation of the K-N cross-section. 
In the shock rest frame, 
the direction of the flow velocity is negative along the $z$-direction as shown in Fig.~\ref{fig:sokudo}; 
most of photons are then scattered to the downstream,
and the peak of the directional angle appears in the backward along the $z$-direction. 

Now, the relation between the mean free path and photon traveling distance during $\Delta t$ is expressed as follows: 
\begin{equation}
\label{eq:alpha}
\alpha = c \left( \frac{\Delta t}{\Delta s_{min}} \right),
\end{equation}
where $\Delta s_{min}$ is the mean free path of a photon traveling opposite to the flow-velocity direction. 
Here, the free path of the photon in the CMF is transformed to that in the OBF 
by the same transformation as Eq.~(\ref{eq:trans_free-path}), 
and $\Delta s_{min}$ is the shortest mean-free-path when $\vector{v} \cdot \vector{\Omega} / c = -v/c$. 
We employed $\Delta s_{min}$ for obtaining the strictest constraint for $\Delta t$. 
This prescription is set from the idea that the optical depth is invariant in any inertial frame.
The free path decreases as the shock speed increases due to the Lorentz contraction, 
and the required $\Delta t$ is shortened.

\begin{figure}[t]
\begin{center}
  \includegraphics[width=1.0\linewidth]{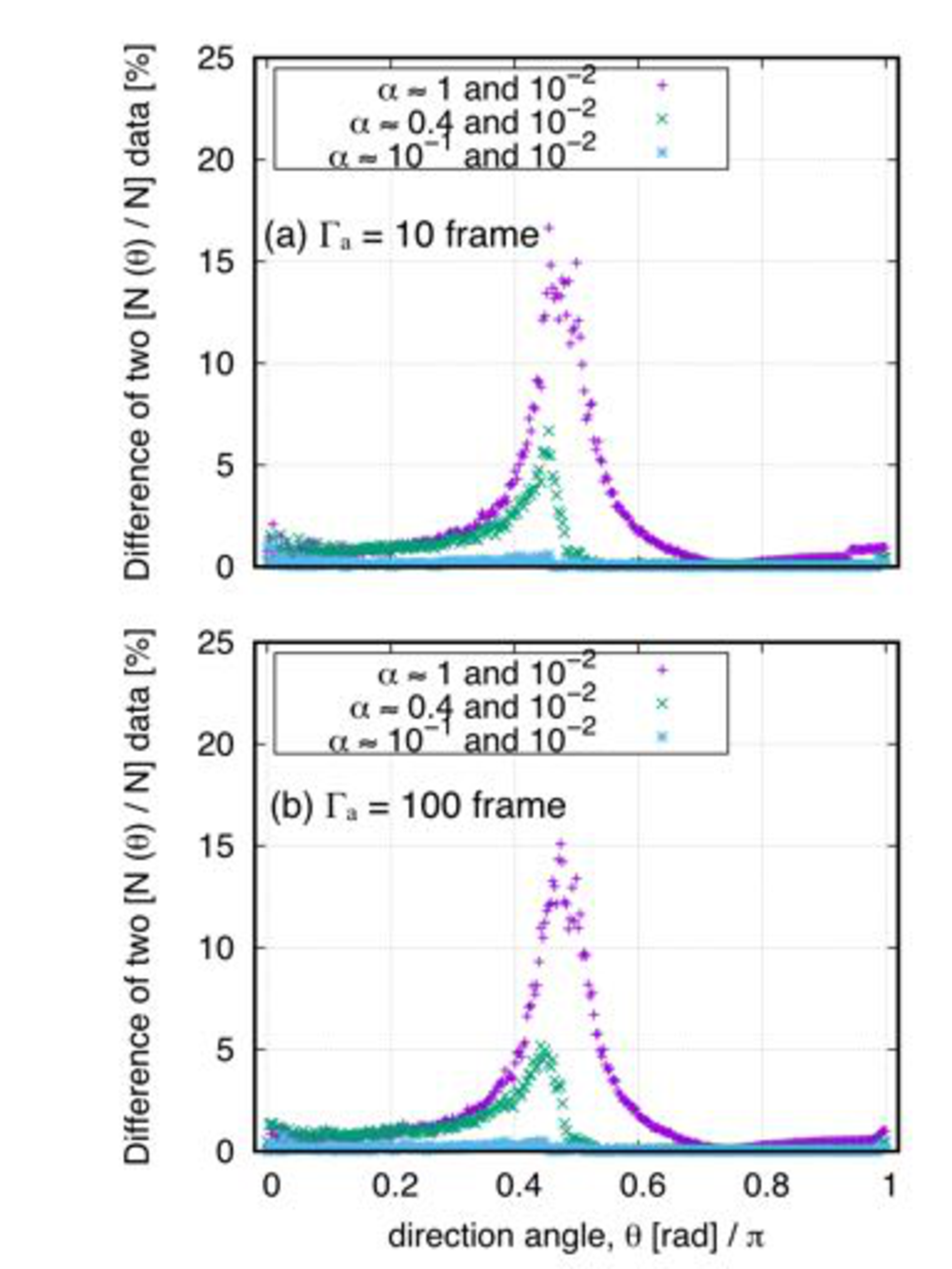}
  \vspace{-10 mm}
  \caption{Relative differences between two distributions with different $\alpha$ values. (a) $\Gamma_a = 10$ frame and (b) $\Gamma_a = 100$ frame.}
  \label{fig:direc_dthenka_differ_010-100}
\end{center}
\end{figure}

In Fig.~\ref{fig:direc_dt_010-100}~(a), 
the result with $\alpha \sim 1$ is significantly different from that with $\alpha \sim 0.4$, 
and the result with $\alpha \sim 0.4$ is slightly different from that with $\alpha \sim 10^{-1}$. 
On the other hand, the result with $\alpha \sim 10^{-1}$ is in good agreement with that with $\alpha \sim 10^{-2}$. 
Thus, the result of the angular distribution in the $\Gamma_a = 10$ frame is converged at $\alpha \sim 10^{-1}$.
The relative differences between two distributions with different $\alpha$ values in the $\Gamma_a = 10$ frame 
are shown in Fig.~\ref{fig:direc_dthenka_differ_010-100} (a).
Here, the angular distribution with $\alpha \sim 10^{-2}$ is regarded as a sufficiently converged one. 
The relative difference between the result with $\alpha \sim 1$ and that with $\alpha \sim 10^{-2}$ is significant 
and the relative difference between the result with $\alpha \sim 0.4$ and that with $\alpha \sim 10^{-2}$ is smaller, 
while the relative difference between the result with $\alpha \sim 10^{-1}$ and that with $\alpha \sim 10^{-2}$ are almost negligible.
The result is significantly different between $\alpha \sim 1$ and $10^{-1}$ 
because the condition of $\alpha \sim 1$ implies that the mean free path is almost the same 
as the photon traveling distance during $\Delta t$. 
Hence, photons may not be scattered before updating the flow-field due to insufficiently small $\Delta t$.
That is, $\alpha \sim 1$ can be regarded as a threshold condition on whether photons are scattered or not scattered 
before the flow-field update. 
Indeed, with $\alpha$ smaller than unity,
the result approaches to the converged one as shown in Fig.~\ref{fig:direc_dt_010-100}.
Here, $\Delta t$ corresponding to $\alpha \sim 10^{-1}$ is $\sim$$10^{-3}$ s in the $\Gamma_a = 10$ frame. 
The time step for photon transport is comparable to or smaller than $\Delta t$ for updating the flow-field.

Similarly, in the $\Gamma_a = 100$ frame, 
the result is converged at $\alpha \sim 10^{-1}$, as shown in Fig.~\ref{fig:direc_dt_010-100}~(b). 
Here, $\Delta t$ corresponding to $\alpha \sim 10^{-1}$ is $\sim$$10^{-4}$ s in the $\Gamma_a = 100$ frame.
This implies that the applicable $\Delta t$ in the $\Gamma_a =100$ frame is smaller by 
one order of magnitude 
than that in the $\Gamma_a = 10$ frame 
because $\Delta s_{min}$ in the $\Gamma_a = 100$ frame is smaller by 
one order of magnitude 
compared to that 
in the $\Gamma_a = 10$ frame as a result of the Lorentz contraction. 
The angular distribution results in Fig.~\ref{fig:direc_dt_010-100} were 
computed in two different inertial frames; 
however, these were transformed to the shock rest frame, 
so that similar features were obtained.
The differences between two distributions with different $\alpha$ values in the $\Gamma_a = 100$ frame 
are also shown in Fig.~\ref{fig:direc_dthenka_differ_010-100}~(b), 
and the difference between the result with $\alpha \sim 10^{-1}$ and that with $\alpha \sim 10^{-2}$ is 
smaller than that of the other two cases and is almost negligible. 

\begin{figure}[t]
\begin{center}
  \includegraphics[width=0.8\linewidth]{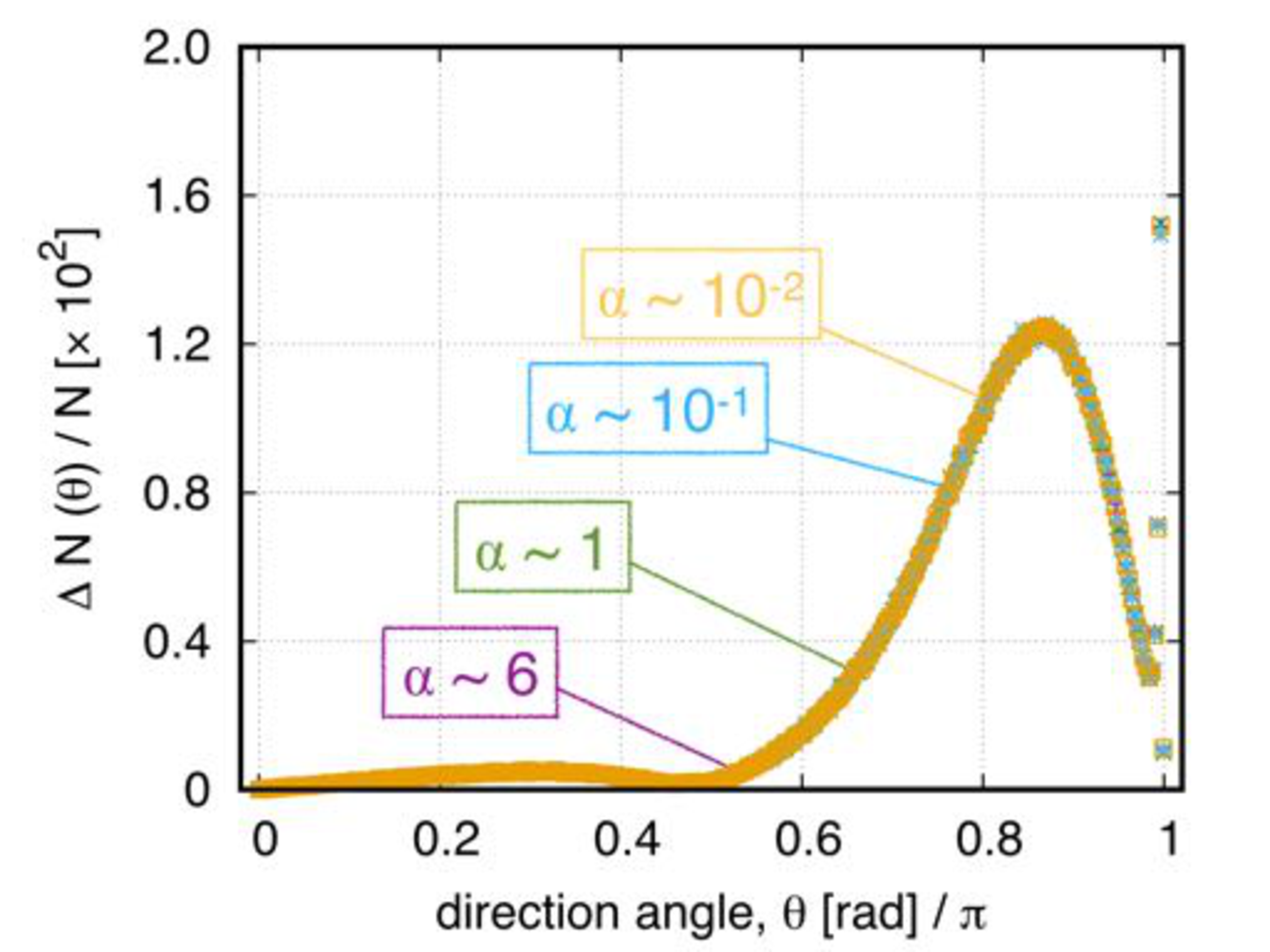}
  \caption{Angular distributions with various $\Delta t$ values in the shock rest frame.}
  \label{fig:direc_dt_001}
\end{center}
\end{figure}

These differences with different $\Delta t$ values are interpreted as follows. 
The simulation result is never dependent on $\Delta t$ in the shock rest frame, as shown in Fig.~\ref{fig:direc_dt_001}. 
In the shock moving frames, however, 
photons can escape from the computational domain or overtake the shock wave front 
before scattering at incorrect timing when $\Delta t$ is not sufficiently small 
because the downstream-side and upstream-side computational boundaries 
and the cell interface (i.e., the shock front) move with the shock speed. 
As the difference between the boundary moving velocity 
(that is, shock front moving velocity) and the speed of light along the $z$-axis is small 
because the optical depth is not so large, 
the large $\Delta t$ fails to produce accurate simulation results. 
The computational grid catches up with the photons or the photons catch up with the grid 
before the photons complete their free path due to a large $\Delta t$; 
consequently, the angular distribution varies according to the value of $\Delta t$. 

In both the $\Gamma_a = 10$ and $100$ frames, with a suitable time interval, 
the value of $\alpha$ is less than $\sim$$10^{-1}$. 
Therefore, we should adopt $\Delta t$ that resolves the mean free path to almost ten steps.
Now, the value of $\alpha \sim 10^{-1}$ is not a critical value, but rather a sufficient value, 
because we change $\Delta t$ by 
one order of magnitude 
in the range of $\alpha \sim 1$ -- $10^{-1}$ 
for examining whether the angular distribution is converged. 
Examination with a smaller change of $\Delta t$ in the range from $10^{-1}$ to $1$ is required 
for determining the critical value of $\alpha$.

\subsection{Comparison of spectra in different inertial frames}
\begin{figure}[t]
\begin{center}
  \includegraphics[width=0.8\linewidth]{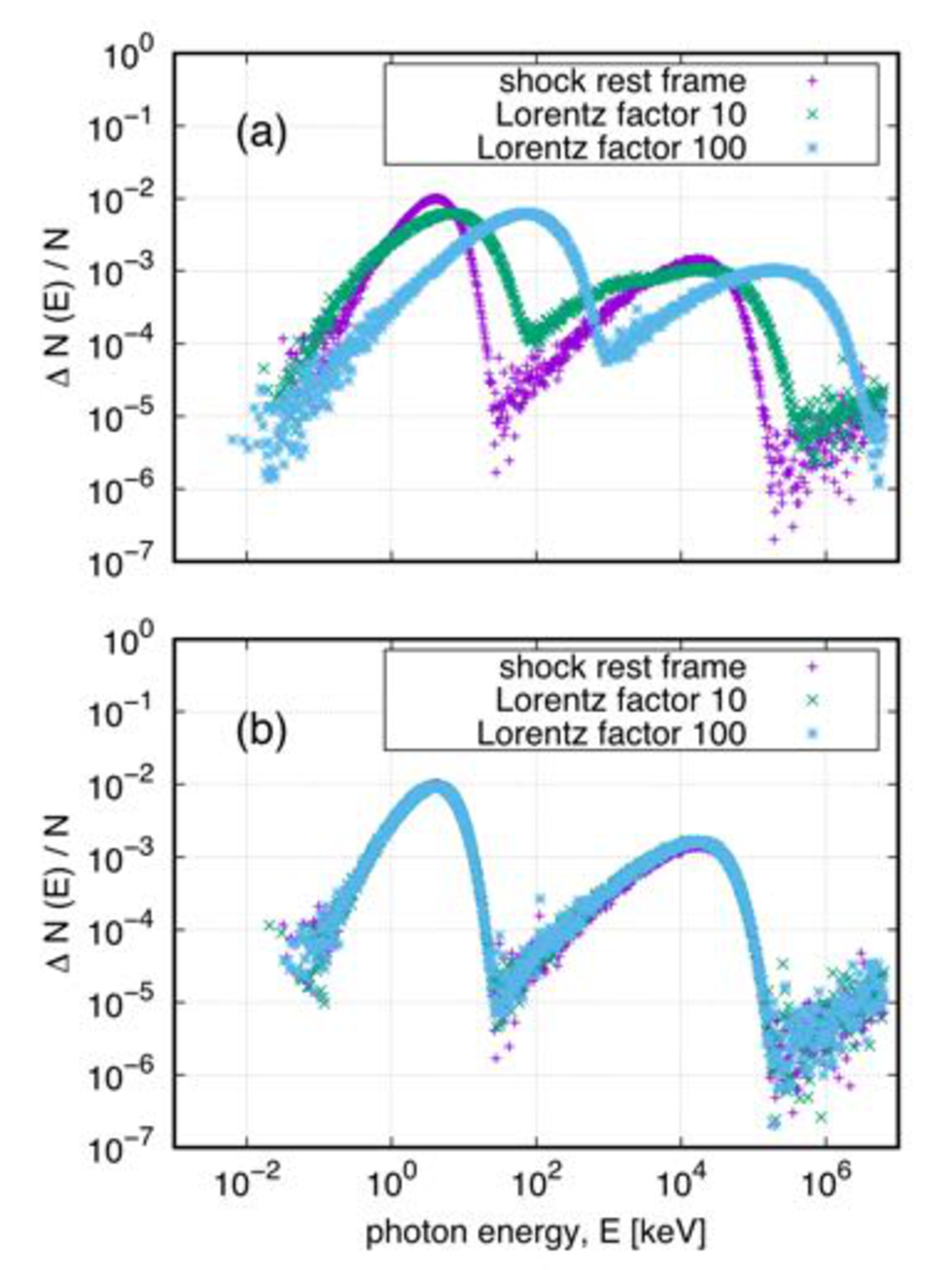}
  \caption{(a) Spectra in each inertial frame and (b) spectra after transforming to the shock rest frame.}
  \label{fig:spec_each-trans}
\end{center}
\end{figure}

The energy spectra of photons escaped from the computational domain are shown in Fig.~\ref{fig:spec_each-trans}~(a) 
in each inertial frame with $10^{6}$ sample particles. 
Here, $\Delta t$'s in the shock moving frames are set as $\alpha \sim 10^{-1}$, as in the earlier section. 
The flow velocity at the downstream side of the shock wave increases along the $z$-axis 
as the apparent shock-Lorentz-factor $\Gamma_a$ increases, 
as shown in Fig.~\ref{fig:sokudo}. 
Consequently, the peak energies of the emission spectra are shifted to the high-energy side by the Doppler effect.
The energy spectra transformed from each frame to the shock rest frame are shown in Fig.~\ref{fig:spec_each-trans}~(b).
The spectra are identical among the three frames after transformation, 
and they are converged with $\Delta t$ satisfying $\alpha \sim 10^{-1}$.
The second peak in the high-energy side is formed through bulk Compton scattering, 
in which the photon energy is boosted by collisions with relativistic electrons. 
The energy of the scattered photon is higher by a square of the flow Lorentz factor $\Gamma^2$ 
compared to that of the incident photon \cite{rybicki}. 
Here, the Lorentz factor of the flow velocity in the upstream side of the shock wave 
in the shock rest frame, $\Gamma_{s2}$, is $\sim$$100$. 
Since the first peak energy is at a few times of $10^0$ keV in Fig.~\ref{fig:spec_each-trans}~(b), 
the second peak energy must be located at a few times of $10^0\ {\rm keV} \times \Gamma_{s2}^2$, 
that is, a few times of $10^4$ keV, 
which is in good agreement with the second peak position in Fig.~\ref{fig:spec_each-trans}~(b).
GeV-order photons are also found in the spectrum because the K-N cross-section and the energy loss by the Compton scattering 
is not included here and energy cut-off does not appear.

Note that we pick photons that escaped from all the computational boundaries, 
not only the upstream side of the shock wave but also the downstream side and the boundary parallel to the cylinder axis. 
As we identify whether the simulation results are in good agreement 
among all the different frames, 
all photons treated in the computation should be sampled.
In fact, if the observer is at a certain position, 
photons traveling in the direction opposite to the observer can never be observed. 
Only photons that enter the view angle of the stationary observer should be sampled 
to obtain the spectra measured in the observations; 
however, the number of sampled photons decreases, 
and large statistical errors may appear in the MC computation.
We discuss about this issues in \ref{sec:convergence_test}.

\subsection{Effect of shock-wave position}
\label{sec:shock_position}
So far, we have considered that the shock wave does not move during $\Delta t$. 
The shock speed is assumed to be constant. 
Thus, in this section, when photons travel across the shock front during $\Delta t$, 
the true position of the shock front can be calculated analytically with $v_s t$. 
Therefore, we can correctly determine whether photons can overtake the shock wave 
since we can compute the true shock position. 
Similarly, we can correctly determine 
whether photons escape from the computational boundary 
because we can compute the true boundary position 
(which moves at the same speed as the shock wave). 
This additional technique is required for calculating the scattering process with the true shock position; 
thus, a more accurate energy spectrum can be obtained with this technique. 

\begin{figure}[t]
\begin{center}
  \includegraphics[width=0.8\linewidth]{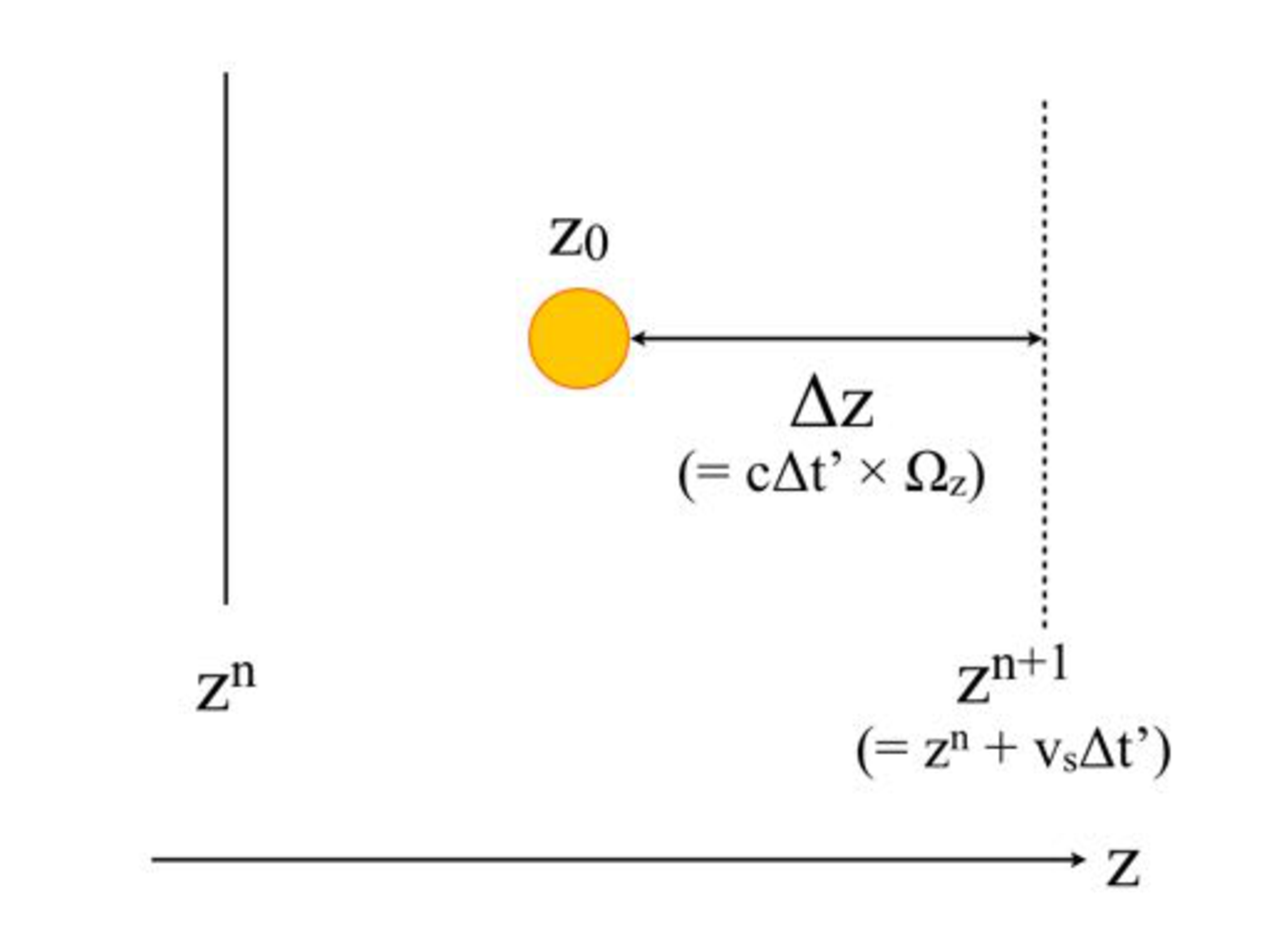}
  \caption{Positions of a photon and computational grids.}
  \label{fig:ph-grid}
\end{center}
\end{figure}

Let us assume that a photon, initially located at a position $z_0$, catches up with the grid 
(which moves at a uniform speed $v_s$) after a time duration $\Delta t' < \Delta t$.
The relation between the positions of the photon and of the moving grid are shown in Fig.~\ref{fig:ph-grid}, 
where $z^n$ denotes the position of the grid (cell interface) at an initial time. 
After a time interval $\Delta t'$, the interface $z^n$ moves to a new location $z^{n+1} = z^n + v_s \Delta t'$.
The distance that the photon travels in the $z$-direction during $\Delta t'$ can be computed 
as $\Delta z = c \Delta t' \Omega_z$, where $\Omega_z$ is the $z$-component of the photon directional unit vector. 

We split a time interval $\Delta t'$ as $\Delta t' = \Delta t'' + \varepsilon$, 
where $\varepsilon << \Delta t''$.
After a photon travels during a time interval $\Delta t''$, 
it stops before the cell interface; 
otherwise, the mean free path cannot be accurately computed. 
This is because the local conditions along the photon path (e.g., the density or the flow velocity) 
might not be known if the photon overtakes the cell interface. 
This procedure prevents the artificial escape of photons at incorrect time from the computational domain 
and avoids that photons overtake the shock front at false locations. 

\begin{figure}[t]
\begin{center}
  \includegraphics[width=0.8\linewidth]{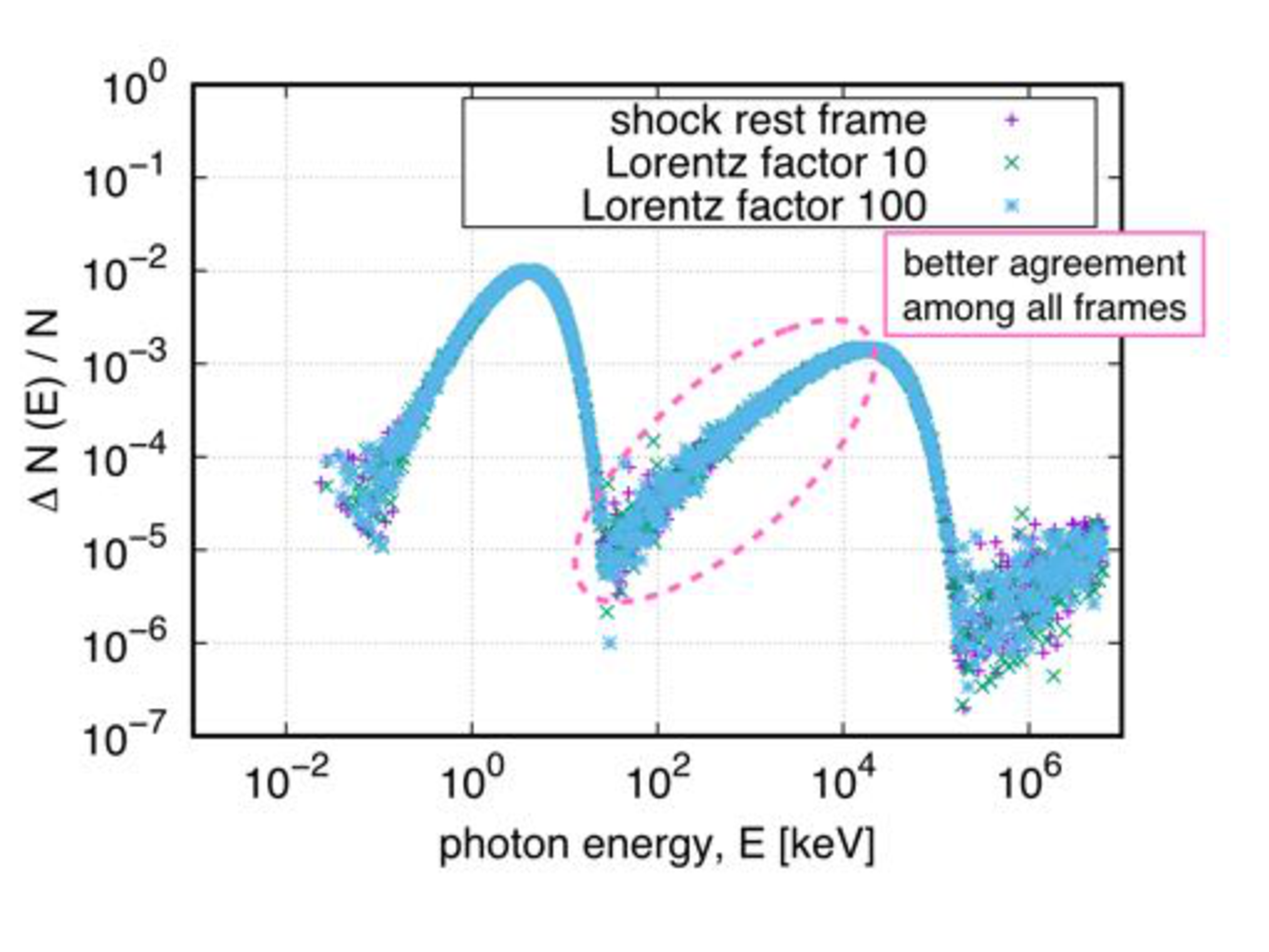}
  \caption{Spectra after transforming to the shock rest frame obtained by analytically setting the true position of the shock front.}
  \label{fig:spec_trans_impli}
\end{center}
\end{figure}

The spectra obtained after transforming to the shock rest frame are shown 
in Fig.~\ref{fig:spec_trans_impli}. 
We used $10^{6}$ sample particles, 
and $\Delta t$ values in the shock moving frames are set to satisfy $\alpha \sim 10^{-1}$ as in the previous section. 
The energy distributions in the second peak of the high-energy side ($100\ {\rm keV} - 10\ {\rm MeV}$) 
computed in the three different frames are in much better agreement 
than those displayed in Fig.~\ref{fig:spec_each-trans} (b).
In a coupled computation of radiative transfer with hydrodynamics, 
it is not possible to accurately calculate the position of the shock front 
since the shock speed may develop with time.
Thus, the time-step size $\Delta t$ in coupled computation should be set sufficiently small to minimize the errors 
in determining the location of the photon when it crosses the shock front. 

Here, since the constraint in the time step is caused by the shock displacement, 
it should not be considered in the shock rest frame because the position of the shock front is always fixed.
Indeed, the comparison among the different $\alpha$'s in the shock rest frame is shown in Fig.~\ref{fig:direc_dt_001}, 
and the angular distributions with any $\alpha$'s are in good agreement each other.

\subsection{Effect of Compton scattering}
\begin{figure}[t]
\begin{center}
  \includegraphics[width=0.8\linewidth]{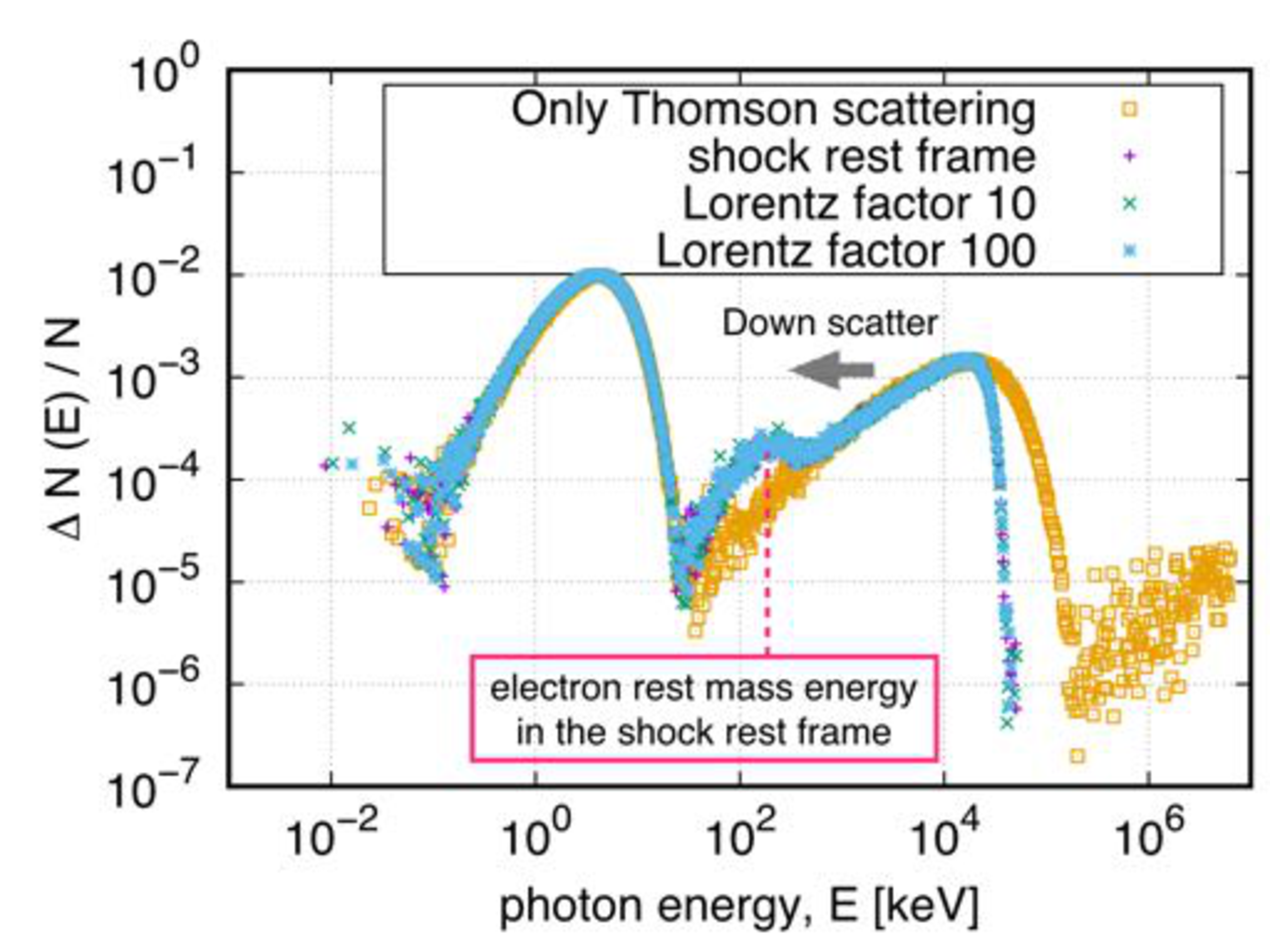}
  \caption{Spectra after transforming to the shock rest frame with Compton scattering.}
  \label{fig:spec_trans_comp}
  \includegraphics[width=0.8\linewidth]{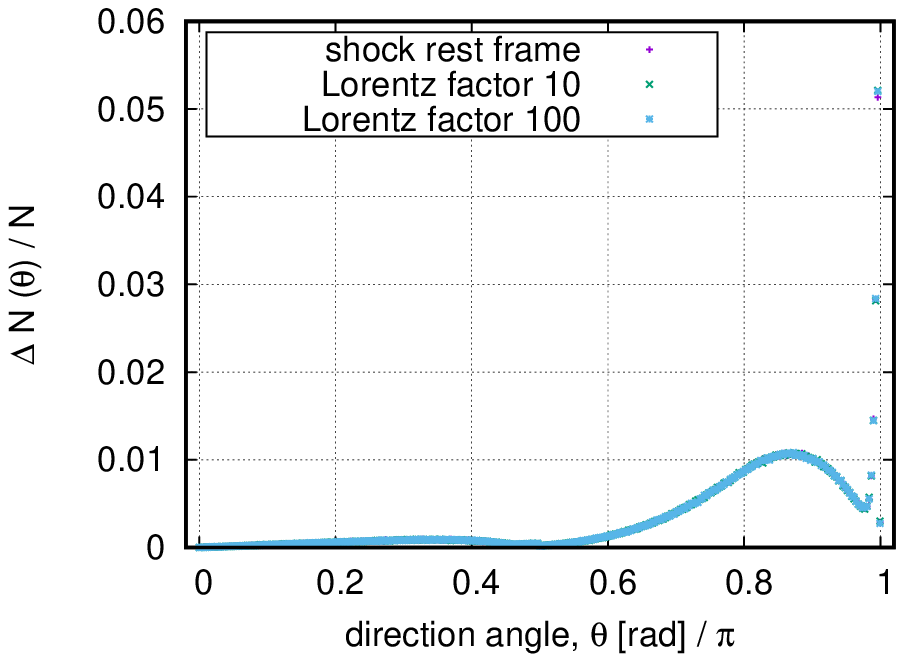}
  \vspace{10 mm}
  \caption{Angular distribution after transforming to the shock rest frame with Compton scattering.}
  \label{fig:direc_trans_comp}
\end{center}
\end{figure}

Here, we also examined the computation with Compton scattering 
by employing the K-N cross-section and considering energy loss 
for the photons with energy greater than 10 keV, 
and the cross-section for Thomson scattering was employed for the other photons. 
As in the earlier sections, we used $10^{6}$ sample particles. 
The time interval $\Delta t$ in the shock moving frames are set as satisfying $\alpha \sim 10^{-1}$, 
and the true position of the shock front is calculated 
when a photon overtakes the shock wave, as in the Sec. \ref{sec:shock_position}.
The spectrum after transforming to the shock rest frame with Compton scattering is shown 
in Fig.~\ref{fig:spec_trans_comp}.
The simulation results of all the frames agree each other even with Compton scattering.
The cut-off appears in the high-energy side, 
in contrast to the case in which only Thomson scattering is considered,
because the high-energy photon loses its energy through the Compton scattering process. 
An increase in the number of photons 
in the range of a few times of $10^{2}\ {\rm keV}$ 
results from the shift of the down-scattered photon losing its energy down to the electron rest-mass energy.
The readers can refer to \ref{comp_analysis} for the details of the peak found at the electron rest-mass energy.

\begin{figure}[t]
  \begin{center}
    \vspace{-10mm}
    \includegraphics[width=0.8\linewidth]{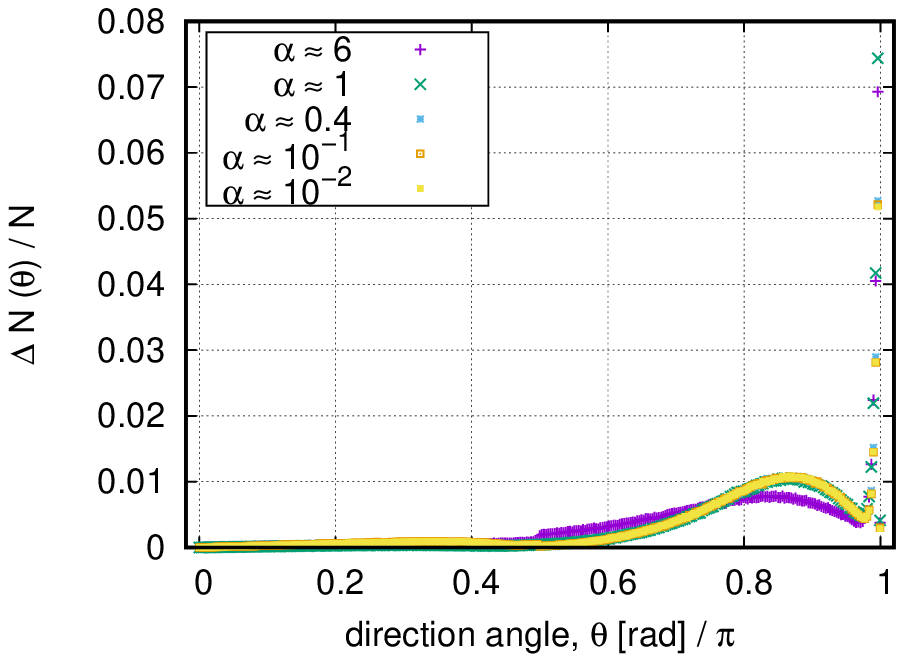}
  \vspace{10 mm}
  \caption{Angular distributions after transforming to the shock rest frame with various $\Delta t$ values with Compton scattering ($\Gamma_a = 10$).}
  \label{fig:direc_dthenka_comp}
  \includegraphics[width=0.8\linewidth]{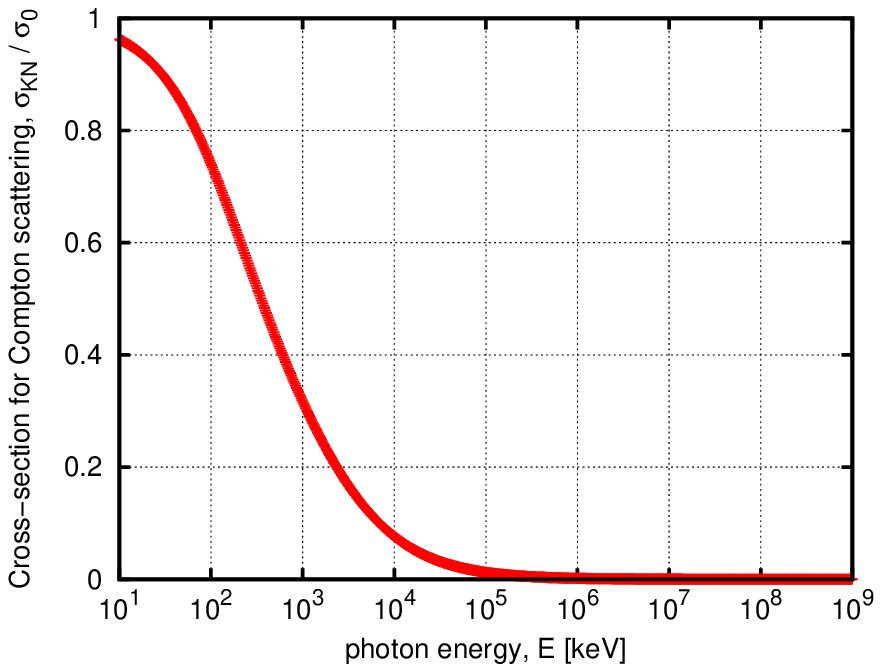}
  \vspace{10 mm}
  \caption{Variation of the K-N cross-section depending on the photon energy.}
  \label{fig:kn-cross}
\end{center}
\end{figure}

The angular distribution with Compton scattering transformed to the shock rest frame 
is shown in Fig.~\ref{fig:direc_trans_comp}.
The property that a peak appears in the backward direction at $\theta / \pi \sim 0.9$ along the $z$-axis 
is similar to that in the case in which only Thomson scattering is considered.
The angular distributions after transforming to the shock rest frame with various $\Delta t$ values 
are shown in Fig.~\ref{fig:direc_dthenka_comp}. 
The result with $\alpha \sim 6$ is significantly different from the others, 
and the results with $\alpha \sim 1,\ 0.4,\ 10^{-1},$ and $10^{-2}$ are in good agreement each other.
This means that the angular distribution is converged for $\alpha \lesssim 1$, 
and the value of $\alpha$ to obtain a converged result is larger by about one order of magnitude 
than in the case in which only Thomson scattering is included. 
This happens because the effect of the K-N cross-section acting in the case in which Compton scattering is considered. 
The variation in the ratio of the total K-N cross-section, $\sigma_{KN}$, 
to the total cross-section for Thomson scattering, $\sigma_{0}$, 
depending on the photon energy is shown in Fig.~\ref{fig:kn-cross}.
The cross-section decreases as the photon energy increases. 
Therefore, the high-energy photons are not scattered as frequent as their low-energy counterparts.
Thus, the mean free path in the case in which Compton scattering is included is larger than that with only Thomson scattering. 
Consistently, the value of $\Delta t$ needed for obtaining a converged angular distribution 
is larger than that with only Thomson scattering (see Eq.~(\ref{eq:alpha})). 

\begin{figure}[t]
  \begin{center}
    \vspace{-10mm}
    \includegraphics[width=0.8\linewidth]{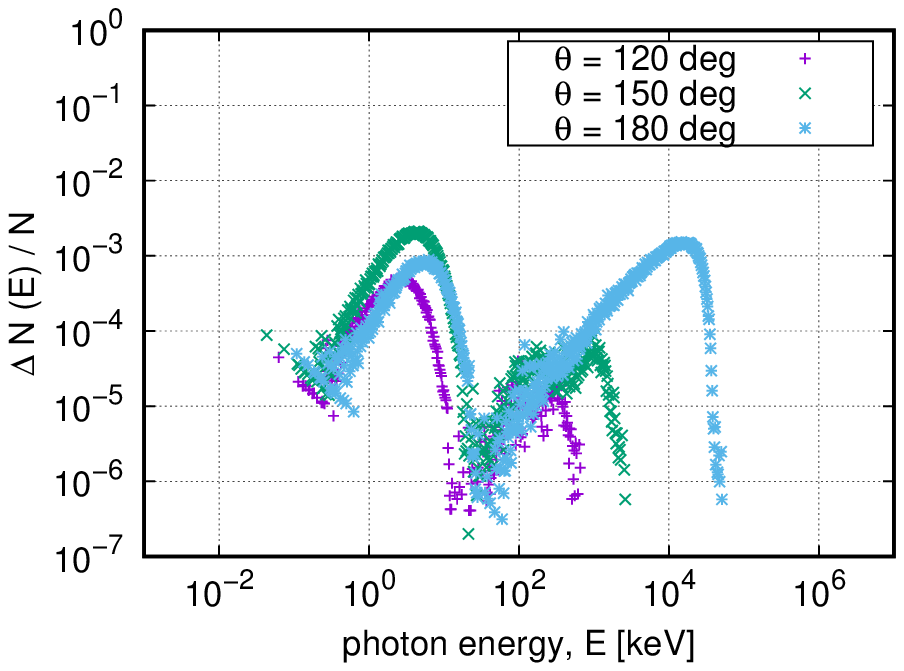}
    \vspace{10 mm}
    \caption{Spectra separated in each direction angle.}
    \label{fig:spec_dir}
    \includegraphics[width=0.8\linewidth]{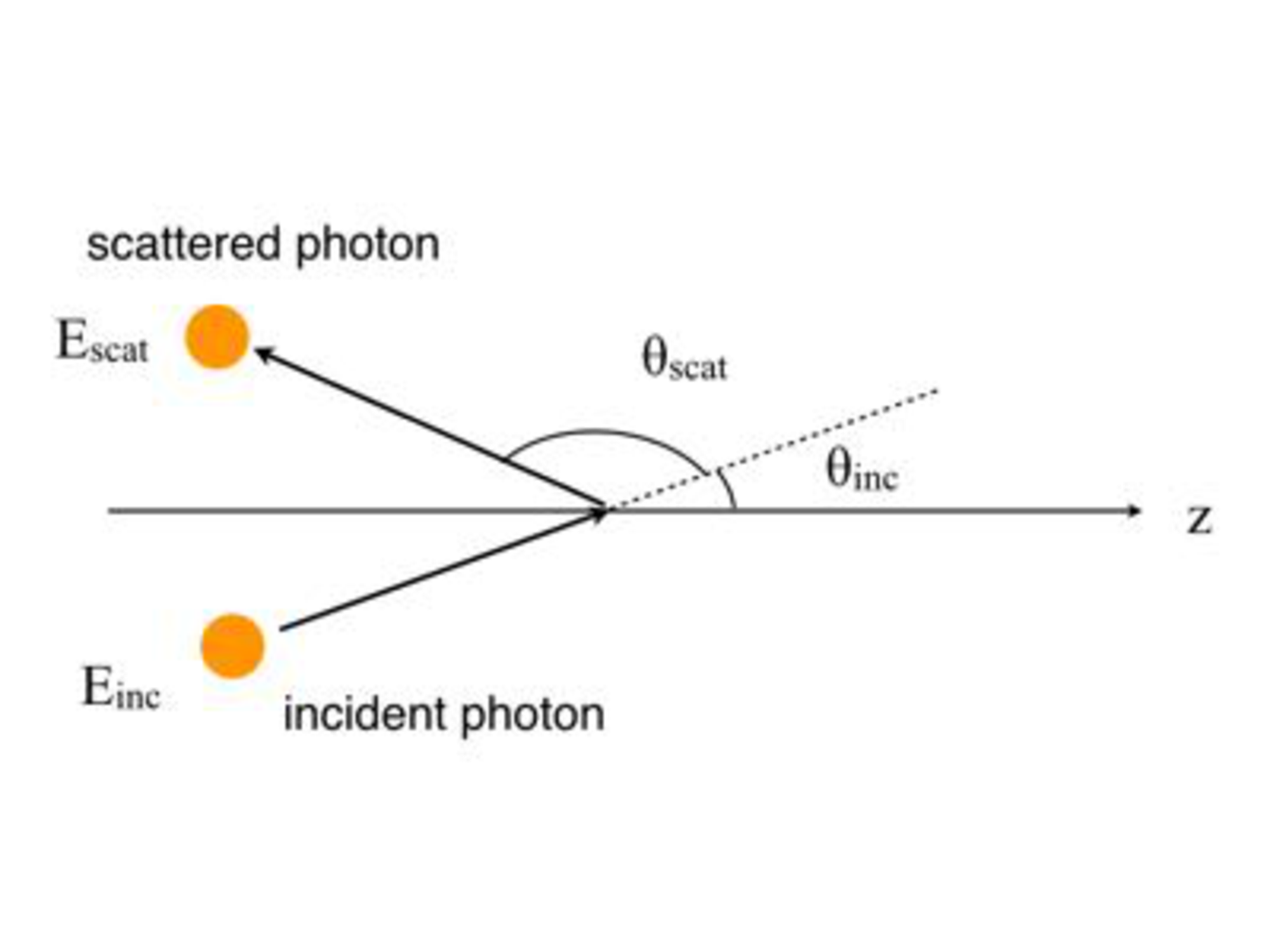}
    \vspace{-10mm}
    \caption{Relation between direction angles of the incident photon and scattered photon.}
    \label{fig:inc_scat_photon}
  \end{center}
\end{figure}

Moreover, the distribution at the direction of $\sim$$180^{\circ}$ with respect to the $z$-axis increases rapidly in Fig.~\ref{fig:kn-cross}.
The flow velocity in the shock rest frame is 
along the negative direction of the $z$-axis, 
and the photons traveling along the direction of the flow 
(in this situation, the angle between the flow velocity and the direction of photon travel is small)  
have high energy, as indicated in Eq.~(\ref{eq:nu_trans}) 
for $\vector{\Omega} \cdot \vector{v}/c \sim 1$.
Therefore, the high-energy photons in the direction of $\sim$$180^{\circ}$ 
with respect to the $z$-axis are not greatly scattered because of the energy dependence in the K-N cross-section, 
and they can escape from the computational domain while maintaining the direction. 
This is why the angular distribution increases rapidly near the direction of $180^{\circ}$.
In order to explore the relation of the direction with the energy of photons, 
the spectra are plotted for photons in the directions of 
$\theta = 120^{\circ}, 150^{\circ},$ and 
$180^{\circ}$ with respect to the $z$-axis, as shown in Fig.~\ref{fig:spec_dir}, 
where a bin of the direction angle is $1^{\circ}$. 
Photons with energy greater than a few MeV do not appear 
in the cases of $\theta = 120^{\circ}$ and $\ 150^{\circ}$, 
while many high-energy photons appear at $\sim$$10$ MeV in the case of $\theta=180^{\circ}$.

\begin{figure}[t]
  \begin{center}
    \includegraphics[width=0.8\linewidth]{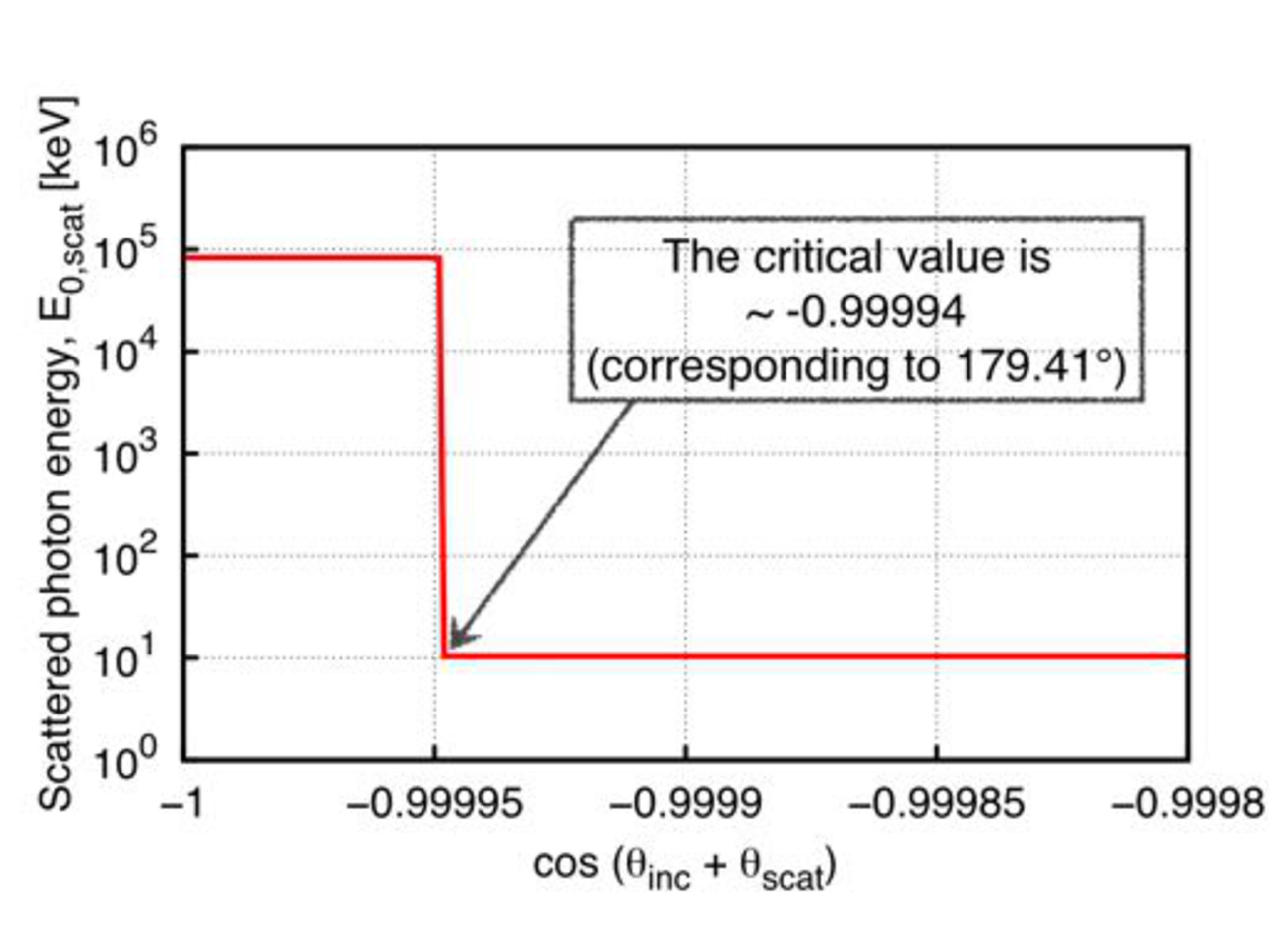}
    \caption{Variation of the scattered photon energy in the CMF versus $\theta_{inc} + \theta_{scat}$.}
    \label{fig:Escat_theta}
    \includegraphics[width=0.8\linewidth]{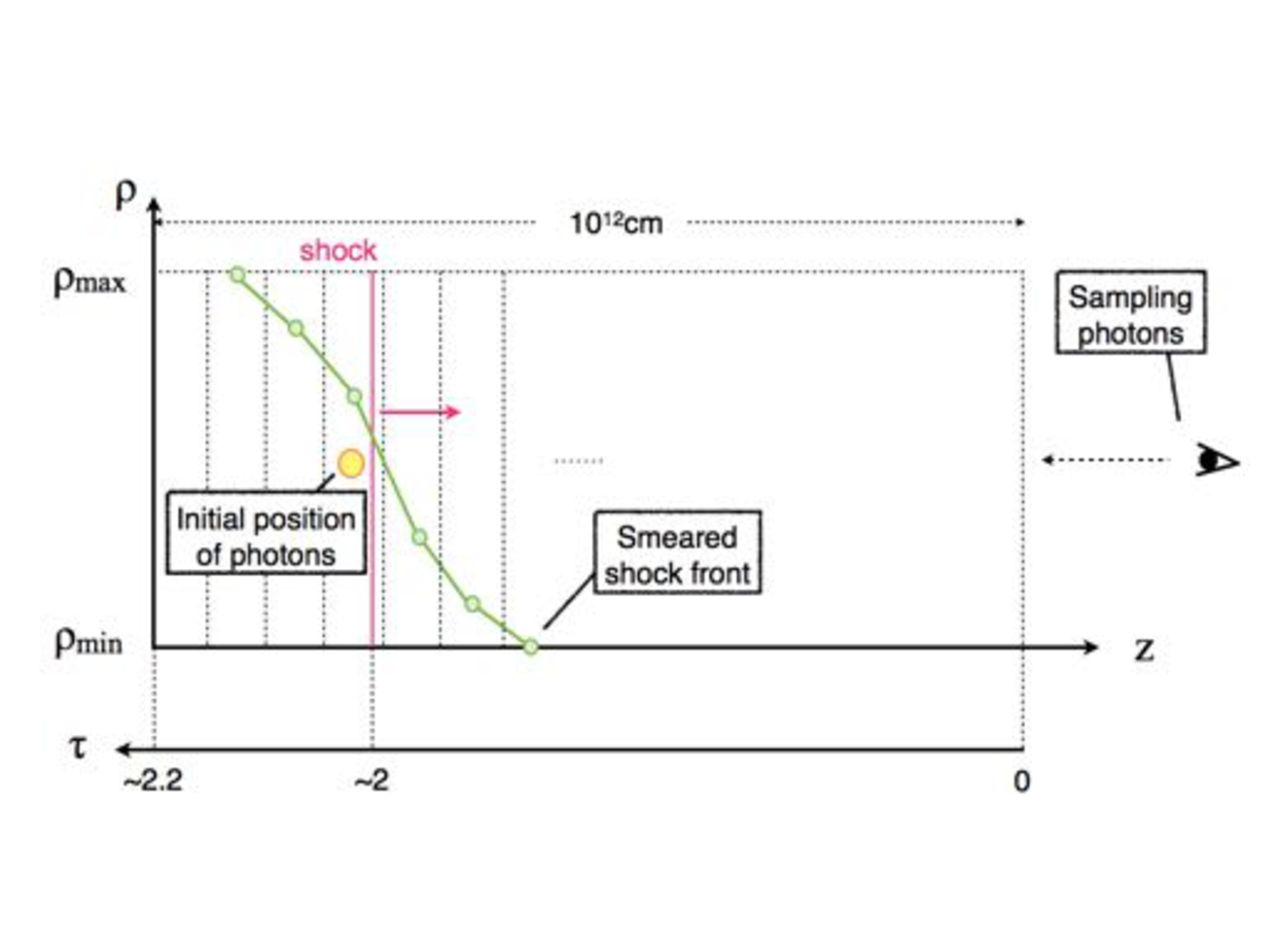}
    \caption{Setting of the smeared shock front.}
    \label{fig:setting_smeared-shock}
  \end{center}
\end{figure}

We also explored the above results through an analytical approach 
to derive Figs.~\ref{fig:kn-cross} and \ref{fig:Escat_theta}
from the analytical formula of the K-N cross-section and Eq.~(\ref{eq:E_scat}). 
The relation between the direction angle of the incident photon, $\theta_{inc}$, 
and scattered photon, $\theta_{scat}$, is illustrated in Fig.~\ref{fig:inc_scat_photon}.
Moreover, the variation of the scattered photon energy in the CMF, $E_{0,scat}$, 
with $\theta_{inc} + \theta_{scat}$ is shown in Fig.~\ref{fig:Escat_theta}.
The energy of the incident photon, $E_{inc}$, is set at the first peak energy 
in Fig.~\ref{fig:spec_trans_comp}, i.e., $\sim$3.6~keV. 
The critical value of $\theta_{inc} + \theta_{scat}$, 
for which the scattered photon energy in the CMF steeply jumps, 
corresponds to $\sim$$179.41^{\circ}$, 
and energy after the jump exceeds $10^4$ keV. 
Photons with direction angles greater than this value are rarely scattered 
since the cross-section of the photon that has energy exceeding $10^4$ keV is close to zero, 
as shown in Fig.~\ref{fig:kn-cross}. 
The critical angle is almost in agreement with the angle at which the angular distribution 
starts to rapidly increase in Fig.~\ref{fig:direc_trans_comp}, 
which is found as $\sim$$178.18^{\circ}$.

\section{Effect of smeared shock front}
\label{sec:smeared_discontinuous}
The effect of the spatial resolution of background flow-field on the radiative transfer computation was discussed in this section. 
The shock front is inevitably smeared in hydrodynamics computation due to numerical diffusion; 
then, we examined the significance of the effect on the emitted spectrum. 
Radiative transfer calculations were implemented for a flow-field with a discontinuous shock wave and 
for one with an artificially smeared shock wave for comparison.

\subsection{Numerical modeling}
The setting of the computational domain for the flow-field with the smeared shock front 
is shown in Fig.~\ref{fig:setting_smeared-shock}, 
where $\tau$ value indicates the initial optical depth.
Computations were executed on a one-dimensional system 
with $10^{5}$ uniform computational cells in the $z$-direction.
The size of the computational domain in the $z$-direction is $10^{12}\ {\rm cm}$, 
which corresponds to $\tau \sim 2.2$. 
All photons are initially placed at a single point immediately behind the shock wave.
The initial position of the shock front corresponds to $\tau \sim 2$. 
Only photons initially emitted in the forward direction along the $z$-axis in the shock rest frame are employed, 
and the others are omitted. 
Radiative transfer computations were implemented on the flow-field with 
both the discontinuous and smeared shock front.
In this computation, 
the inertial frame of interest is the rest frame for the upstream flow. 
Only photons that escape from the forward boundary, 
that is, the photons that can overtake the shock front, were sampled.
The smeared shock front is set in the shock rest frame 
by determining the density distribution as follows:
\[
\hspace{-28mm}
\rho = \left\{  \begin{array}{ll}
    \rho_{max} & (z \le z_{sh}-\delta/2)\\
    \frac{1}{2} \left( \rho_{max} - \rho_{min} \right) 
    \left[1 + {\rm sin} \frac{\pi \left(z_{sh}(t) - z \right)}{\delta} \right]
    + \rho_{min} & (z_{sh} - \delta/2 < z \le z_{sh} + \delta/2) \\
    \rho_{min} & (z > z_{sh}+\delta/2 ),
  \end{array} \right.
\]
\begin{figure}[t]
  \begin{center}
    \includegraphics[width=0.75\linewidth]{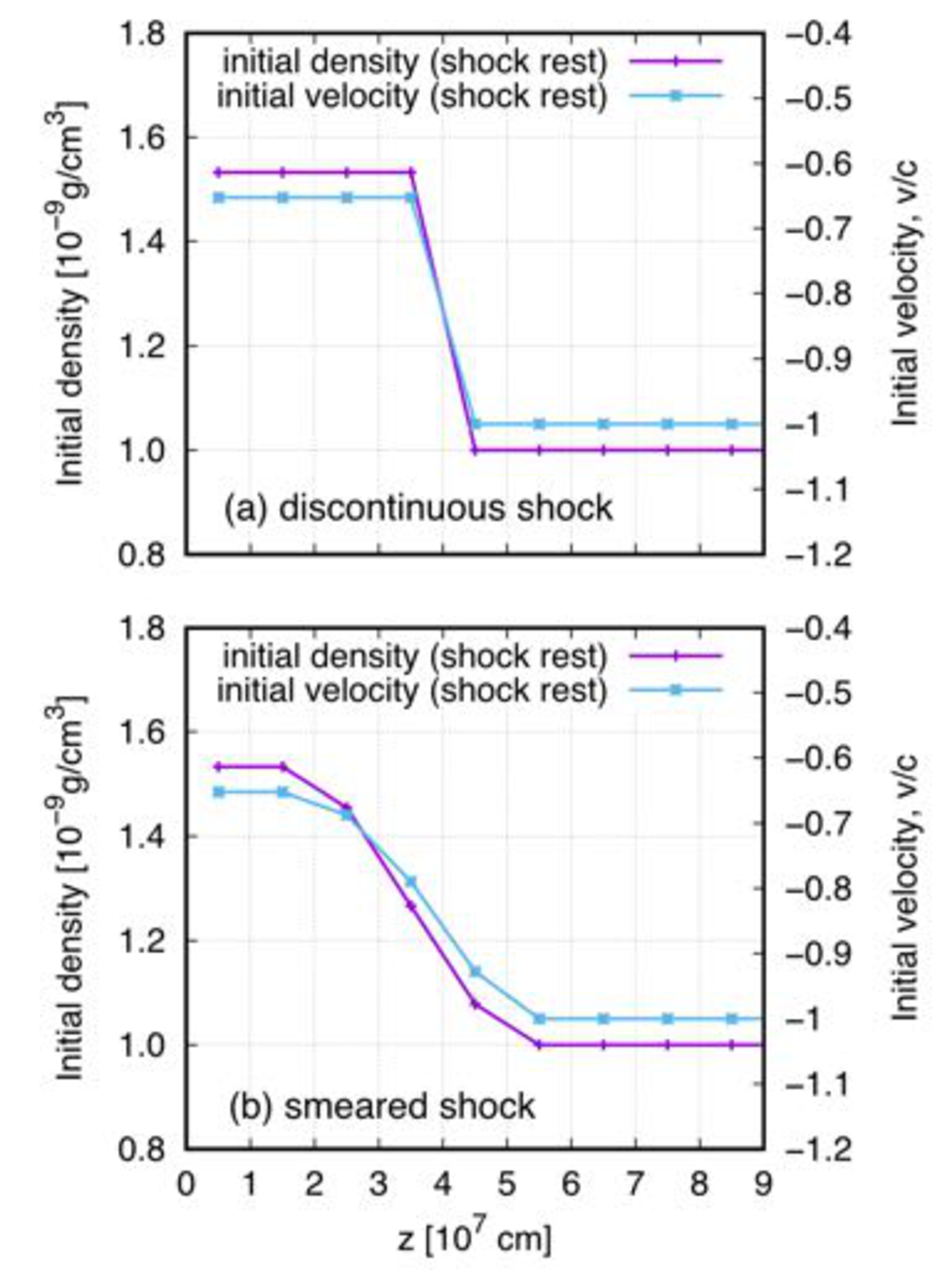}
    \caption{Initial density and flow-velocity distributions in the shock rest frame with (a) the discontinuous shock front and (b) smeared shock front with the shock width $\delta$ corresponding to 4-computational-cell length.}
    \label{fig:ini_dens_velo_d-0-4}
  \end{center}
\end{figure}
where $z_{sh} (t)$ and $\delta$ are the position of the shock wave at a certain time, $t$, 
and shock width, respectively.
The maximum value of the density, $\rho_{max}$, and the minimum value, $\rho_{min}$, 
are associated by the Rankine--Hugoniot relations 
and are set as the downstream and upstream values shown in Table~\ref{tab:butsuriryou}, respectively. 
The distribution of the flow velocity is determined as satisfying 
the equation of continuity, $\left[\rho u^z \right]=0$, 
where $\rho$ is the density in the CMF and $u^z$ is the four-velocity in the shock rest frame.
The equation of continuity can be transformed as follows:
\begin{equation}
\left[ \rho \Gamma_s v^i \right] = 0,
\end{equation}
where $\Gamma_s$ and $v^i$ are the Lorentz factor of the shock velocity 
and the three-velocity, respectively.
Therefore, the equation is simply expressed with the density in the shock rest frame, $\rho_s = \rho \Gamma_s$, 
and the three-velocity, $v^i$, 
without the density transformed to the one in the CMF. 
The shock speed corresponds to $\Gamma_s \sim 100$.
The initial distribution of density and flow velocity in the shock rest frame 
with a discontinuous shock wave are shown in Fig.~\ref{fig:ini_dens_velo_d-0-4}~(a), 
and those with a smeared shock wave are shown in Fig.~\ref{fig:ini_dens_velo_d-0-4}~(b) 
with the shock width $\delta$ corresponding to 4-computational-cell length.

\subsection{Result of spectrum}
\begin{figure}[t]
  \begin{center}
    \includegraphics[width=0.7\linewidth]{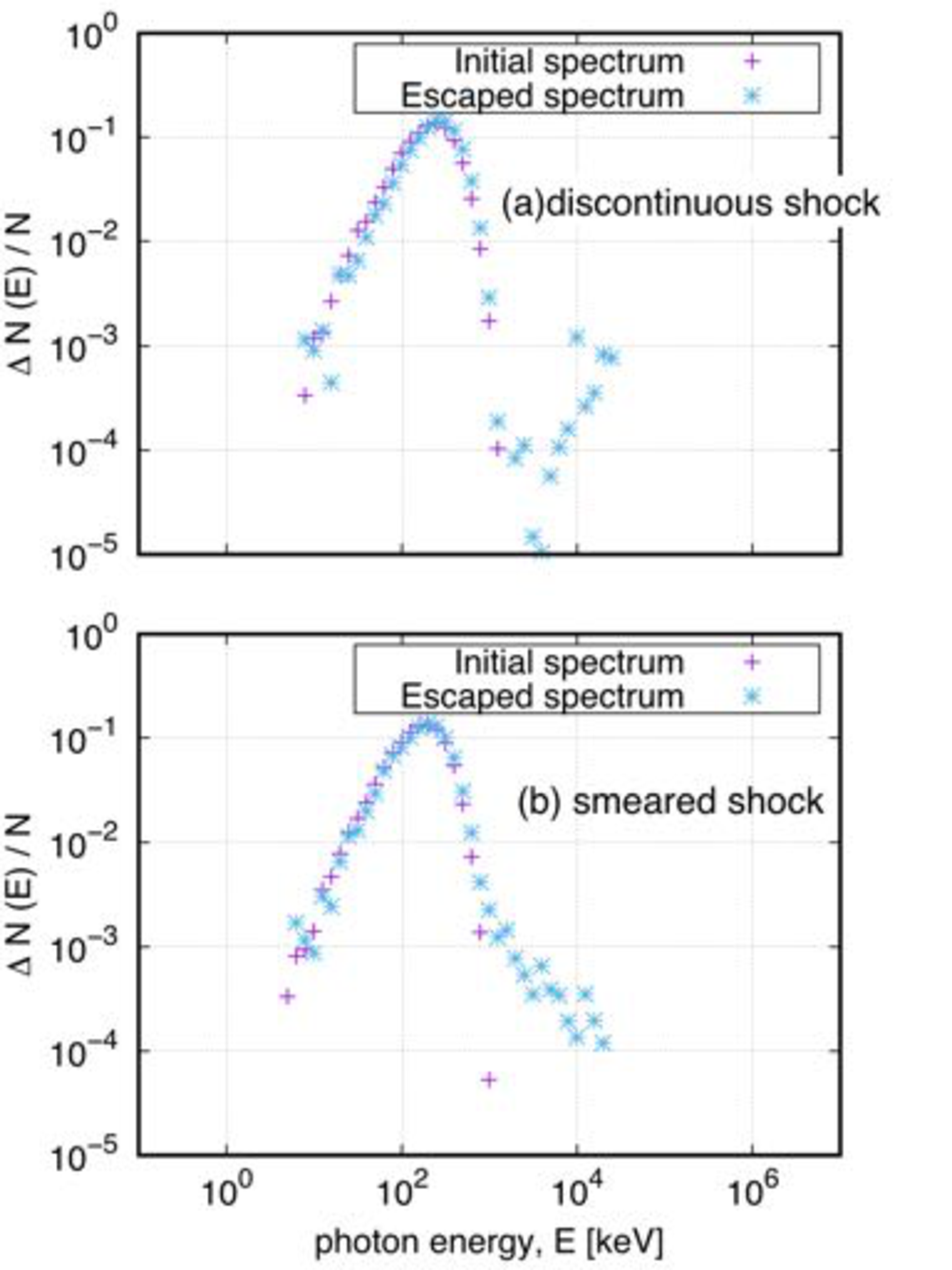}
    \caption{Initial and escaped spectra with (a) the discontinuous shock front and (b) smeared shock front with the shock width $\delta$ corresponding to 4-computational-cell length.}
    \label{fig:spec_euler_d-0-4}
  \end{center}
\end{figure}
We employed $10^{5}$ sample particles 
and set $\Delta t$ as satisfying $\alpha \sim 10^{-1}$ 
in the situation with $\Gamma_a \sim 100$ of the earlier section; 
that is, $\sim$$10^{-4}$ s. 
The spectrum sampling the initially emitted photons and 
the one sampling the photons escaped from the computational domain 
with the discontinuous shock wave are shown in Fig.~\ref{fig:spec_euler_d-0-4} (a), 
and those with the smeared shock wave are shown in Fig. \ref{fig:spec_euler_d-0-4} (b).
The smearing width of the shock wave is set to a 4-cell length.
Both the spectra sampling the escaped photons are different from the initial spectra.
The high-energy photons of $\sim$1--10 MeV order form a power-law curve whose index is positive 
in the case of the discontinuous shock wave, 
while a power-law curve whose index is negative is formed in the case of the smeared shock wave.
High-energy photons over $\sim$10 MeV disappear in both the cases 
since only photons that escape from the forward boundary are sampled.
Photons obtaining high energy via the bulk Compton scattering are scattered to the backward boundary along 
the $z$-axis, and the photon with energy greater than 10 MeV is rarely scattered any more 
since the scattering cross-section is small, as shown in Fig.~\ref{fig:kn-cross}. 
Therefore, high-energy photons escape from the backward boundary 
and cannot contribute to the spectra in these cases.

\begin{figure}[t]
  \begin{center}
    \vspace{-10mm}
    \includegraphics[width=0.8\linewidth]{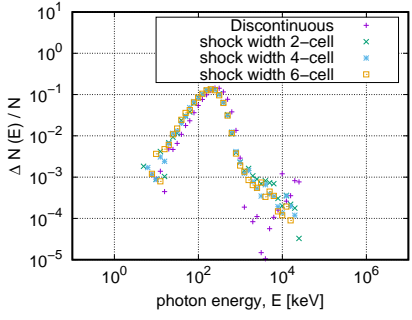}
    \vspace{10 mm}
    \caption{Spectra with discontinuous shock wave and smeared shock waves with three different shock widths.}
    \label{fig:spec_normal-smeared}
  \end{center}
\end{figure}

The spectra with the discontinuous shock wave and smeared shock waves 
with shock widths of 2-cell, 4-cell, and 6-cell length 
are shown in Fig.~\ref{fig:spec_normal-smeared}.
The spectrum in the case with the smeared shock wave has the negative-index power-law in the high-energy side 
and becomes slightly steeper as the shock width increases 
because energy obtained via bulk Compton scattering decreases.
We discuss this effect in the next section.

Actually, it is difficult to use so many computational cells ($10^5$ cells in the computational domain of $10^{12}$ cm) 
in practical multi-dimensional hydrodynamics simulations.
We used such a large number of computational cells to show that the shock structure affects the emitted spectrum 
even when such a fine computational mesh is used. 
If we have 10 or 100 times worse resolution, 
the spectral changes depending on the shock width in the high-energy side may be more remarkable. 

\subsection{Effect of shock width}
\label{sec:shock-width}
\begin{figure}[t]
  \begin{center}
    \includegraphics[width=0.8\linewidth]{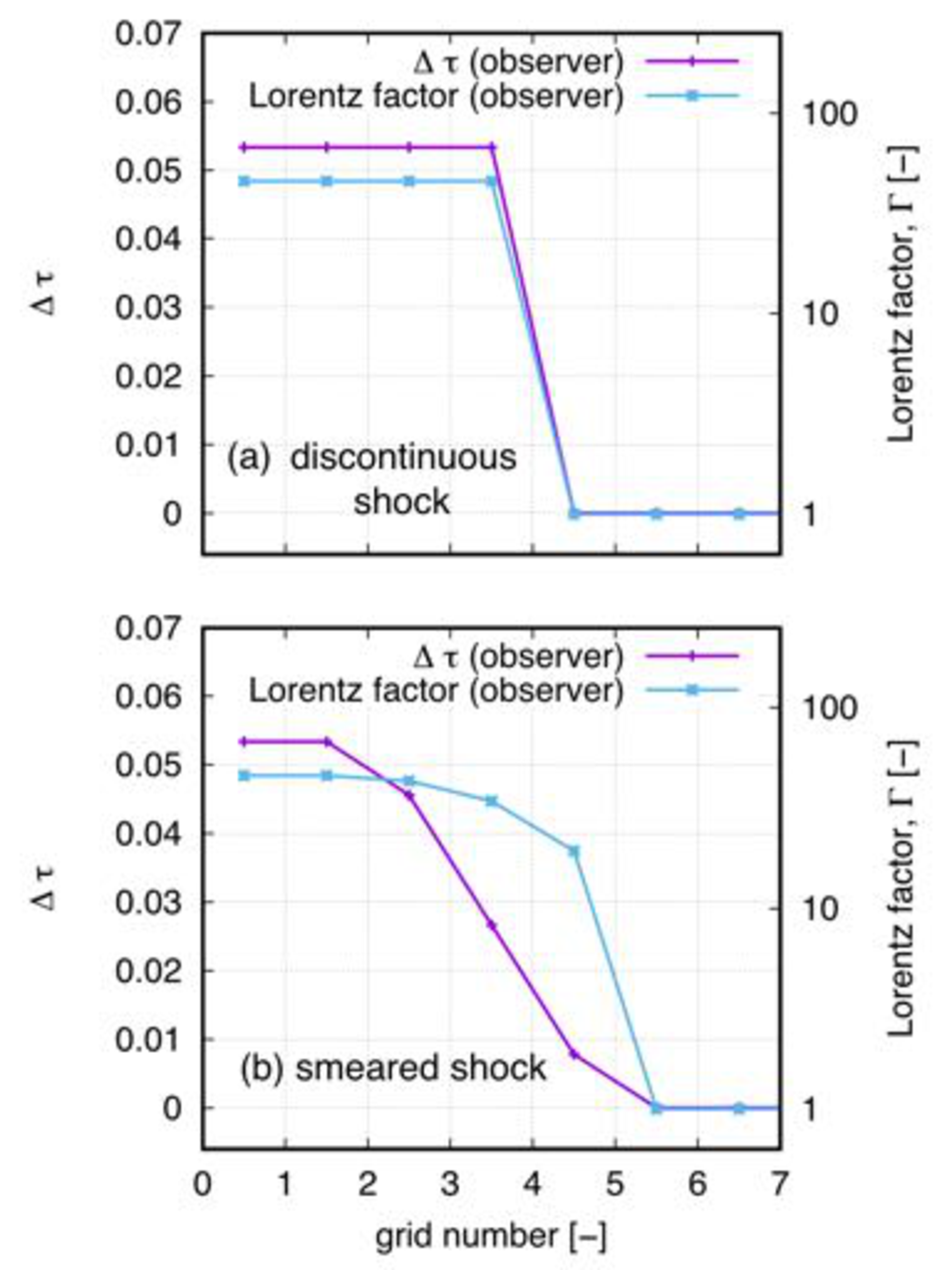}
    \caption{Initial distributions of optical depth and flow velocity with (a) the discontinuous shock front and (b) smeared shock front.}
    \label{fig:tau-velo_d-0-4}
  \end{center}
\end{figure}

We explored the relation between the shape of the spectra and the shock width. 
The initial distributions of optical depth per cell, $\Delta \tau$, 
and the Lorentz factor of the flow velocity in the OBF with the discontinuous shock wave 
are shown in Fig.~\ref{fig:tau-velo_d-0-4} (a),
and those with the smeared shock wave are shown in Fig.~\ref{fig:tau-velo_d-0-4} (b).
The optical depth is an indicator of the probability of scattering, 
and the Lorentz factor is an indicator of energy obtained via bulk Compton scattering.
For obtaining energy via bulk Compton scattering and traveling toward the observer, 
photons should be emitted at the downstream side of the shock wave, 
then scattered at the upstream side to the backward direction, 
and scattered again at the downstream side to the forward direction. 
Therefore, the optical depth is required to be rather large 
around the flow velocity jump.
The optical depth in the case with the discontinuous shock wave is greater than 
that in the case with the smeared shock wave immediately behind the velocity jump.
Photons are rarely scattered at the downstream side in the case with the smeared shock wave, 
in contrast to the case with the discontinuous shock wave; 
consequently, high-energy photons traveling to the forward boundary decrease.

Since the structure of the shock wave can considerably affect
the radiative transfer computation in even the one-dimensional background flow-field, 
we should also consider the effect in the radiation-hydrodynamics coupled computation.
We implemented the coupled computation neglecting the radiation back reaction 
and discussed in \ref{sec:couple-hydro} about the difference of the spectral feature 
from that in the model computations shown in this section.

%

\section{Conclusion}
\label{sec:conclusion}
Radiative transfer computation in the ultra-relativistic flow-field 
was validated as the preliminary study for the coupled computation 
of the radiative transfer and relativistic hydrodynamics 
with the appropriate simulation conditions in the present paper.
We have developed a three-dimensional radiative transfer code 
for the ultra-relativistic background flow-field by using the MC technique, 
and some test calculations were implemented.

We performed the radiative transfer computations in three different inertial frames 
in which the apparent Lorentz factor of the shock speed, $\Gamma_a$, is 1, 10, and 100 
and compared the simulation results in the same frame. 
The angular distributions computed with different $\Delta t$ values showed that 
the simulation result was converged with sufficiently small $\Delta t$. 
The value of $\Delta t$ that resolves the mean free path into ten steps was the sufficient condition 
for obtaining thoroughly converged results.
The value of $\Delta t$ tested in this paper is appropriate for only the case 
without including the radiation back-reaction on the flow-field.
In the case with radiation back-reaction, 
the time step for updating the flow-field will be smaller.
In a fully coupled radiation hydrodynamics simulation, 
energy-momentum is extracted out of hydrodynamic cells and carried away by photons.
Thus, the energy (or pressure) of the cells may become negative 
unless a sufficiently small time step is employed.
In optically thick regions of the flow-field, this restriction can be so tight 
and an implicit or partly-implicit time integration algorithm would be needed. 

The spectra computed in each inertial frame showed that 
the peak energy was shifted to the high-energy side as the flow velocity increases 
because of the Doppler effect.
The spectra transformed from each frame to the shock rest frame were in good agreement,
validating the transformation among the different frames 
in the ultra-relativistic regime.
The spectra were in better agreement on considering the true position 
of the shock front when photons crossed the shock wave.
The spectra with Compton scattering had peak energy at a few times of $10^2\ {\rm keV}$ 
because photons were scattered and lost their energy down to the electron rest mass energy. 
The angular distribution with Compton scattering showed 
a rapid increase along the direction of $\sim$$180^{\circ}$ with respect to the $z$-axis.
The K-N cross-section decreases with the increase in the photon energy, 
and the photons traveling toward the direction of $180^{\circ}$ had high energy; 
then, these photons escaped from the computational domain without scattering any more. 
With Compton scattering, $\Delta t$ for obtaining the converged angular distribution was larger 
than that in the situation with only Thomson scattering 
because of the energy dependence of the K-N cross-section. 
Therefore, the constraint of $\Delta t$ may be relaxed 
in the case that both of Thomson and Compton scatterings are considered. 

We assessed the effect of the spatial resolution of the background flow-field 
on the radiative transfer computation.
Radiative transfer was computed in scenarios 
with a discontinuous and an artificially smeared shock fronts, 
and the simulation results were compared.
High-energy photons with 1--10 MeV order decreased 
in the scenario with the smeared shock front 
compared to that with the discontinuous shock front.
The optical depth immediately behind the shock wave was smaller 
with the smeared shock front compared to that with the discontinuous shock front, 
and photons that obtain high energy through bulk Compton scattering and travel to the observer decreased.
The structure of the shock wave affected the radiation transport even 
in the one-dimensional computation.
In the one-dimensional models presented here, we have employed $10^5$ hydrodynamical cells 
to accurately compute the high-energy tail of the spectrum; 
however, it is difficult to employ so many cells per dimension in multi-dimensional simulation. 
Extending the method proposed here to more than one dimension is handicapped by the huge numerical resolution. 
This is a future work toward the multi-dimensional computation. 

Some test calculations were performed for validating the radiative transfer 
computation on the ultra-relativistic background flow-field 
as the preliminary study for the coupling computation of MC radiation transport with relativistic hydrodynamics.
In the future, we try to combine the developed code in this paper with relativistic hydrodynamics simulation 
to achieve radiative transfer computation 
on the time-dependent ultra-relativistic flow-field 
involving the radiation feedback to the flow-field 
and firstly perform some test computations 
restricted to some regions in the flow-field.

We are grateful to the anonymous referees for their so fruitful comments on this manuscript.
This work was supported by JSPS KAKENHI Grant Number 52638590, 
a Grant-in-Aid for Scientific Research from the Ministry of 
Education, Culture, Sports, Science, and Technology (MEXT) of Japan (24103006, 24740165, and 24244036), 
and the HPCI Strategic Program of the Japanese MEXT. 
This work is partly supported by the Grant-in-Aid for Scientific Research (S:16H06341) and 
the Grant-in-Aid for Young Scientists (B:16K21630) from the MEXT of Japan.
H. N. is supported in part by JSPS Postdoctoral Fellowships for Research Abroad No. 27--348, 
and also supported at Caltech through NSF award No. TCAN AST--1333520.
H. I. is supported in part by the RIKEN pioneering project ‘Interdisciplinary Theoretical Science (iTHES)’.
\\
\\
\\
\\
\appendix

\section{Convergence test for only photons traveling to the observer}
\label{sec:convergence_test}
\begin{figure}[t]
  \begin{center}
    \vspace{-10mm}
    \includegraphics[width=0.8\linewidth]{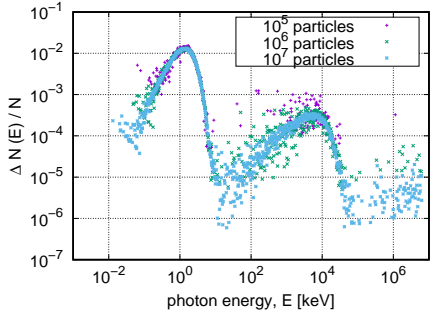}
    \vspace{10 mm}
    \caption{Spectra of escaped photons traveling in the direction to the observer.}
    \label{fig:spec_fore}
  \end{center}
\end{figure}

We computed the spectrum equivalent to Fig.~\ref{fig:spec_each-trans} 
but only for photons that would reach an observer. 
The observer was assumed to be at the upstream side of the shock wave and far enough away from the shock front.
Therefore, we sampled only photons escaped from the forward boundary of the computational domain. 
That is, photons moving at an angle $< 90^{\circ}$ with respect to the shock normal propagation direction were sampled.
The spectrum result has larger statistical errors than that for all photons. 
We compared the result with that for 10 times larger number of photons ($10^7$ particles) 
and smaller number of photons ($10^5$ particles) 
as shown in Fig.~\ref{fig:spec_fore}. 
The spectrum for $10^5$ particles has large statistical errors in the high-energy side, 
and the convex feature is not clear in the range of $10^2 - 10^4$ keV. 
Although the spectrum for $10^6$ particles has somewhat statistical errors, 
the convex feature can be found in the high-energy side. 
The spectrum for $10^7$ particles is better converged. 
Thus, more than $10^7$ particles may be needed for obtaining a realistic observed spectrum.

\section{Energy peak formed by Compton scattering}
\label{comp_analysis}

\begin{figure}[t]
  \begin{center}
    \vspace{-10mm}
    \includegraphics[width=0.8\linewidth]{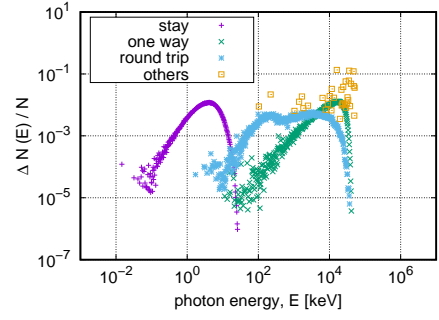}
    \vspace{10 mm}
    \caption{Separated spectra depending on the photon traveling path across the shock wave.}
    \label{fig:spec_scatposi}
  \end{center}
\end{figure}

The peak in the range of $\sim$$10^2\ {\rm keV}$ in Fig.~\ref{fig:spec_trans_comp} is formed 
due to the down-scattered photons from the high-energy side.
The separated spectra depending on the photon traveling path across the shock wave are shown 
in Fig.~\ref{fig:spec_scatposi}.
The word of `stay' in the legend denotes the spectrum for the photons staying in the downstream side 
from emission to escape out from the computational boundary.
Similarly, the word of `one way' denotes the spectrum for the photons traveling 
from the downstream side to the upstream side crossing the shock wave once, 
`round trip' denotes that the photons travel from the downstream side to the upstream side 
and subsequently from the upstream side to the downstream side 
crossing the shock wave twice, 
and `others' denotes the other photons.
The spectrum of `stay' is not significantly different from the initial emitted spectrum; 
however, that of `one way' is shifted to the high-energy side due to bulk Compton scattering.
Moreover, the spectrum of `round trip' has the peak at $\sim$$10^2\ {\rm keV}$ 
since the high-energy photons produced in the upstream side are carried back to the downstream side 
and Compton scattered several times losing their energy down to the electron rest mass energy.

The photon energy after scattering in the CMF, $E_0'$, is calculated as follows: 
\begin{equation}
\label{eq:E_scat}
E_0' = \frac{E_0}{1+ \frac{E_0}{m_e c^2} (1- {\rm cos}\ \Theta)}.
\end{equation}
The second term in the denominator in the right-hand side is predominant 
when the photon energy is larger than the electron rest mass energy, $m_e c^2$; 
consequently, the photon energy after scattering decreases.
In contrast, when the photon energy is smaller than $m_e c^2$, 
this term becomes negligible and the photon energy after scattering 
is not noticeably changed from the incident energy.
Therefore, the photon with high energy loses its own energy through several scattering processes 
down to $\sim$$m_e c^2$, and if its energy falls less than $m_e c^2$, 
the photon energy is unchanged any more.

The flow velocity in the downstream side in the shock rest frame is not highly relativistic, 
so the photon energy corresponding to the electron rest mass $\sim$$0.5\ {\rm MeV}$ in the CMF 
is not so different from that in the shock rest frame. 
Therefore, most of the photons that have high energy after experiencing bulk Compton scattering in the upstream side 
and carried to the downstream side 
hold energy of $\sim$$10^2\ {\rm keV}$.

\begin{figure}[t]
  \begin{center}
    \includegraphics[width=0.8\linewidth]{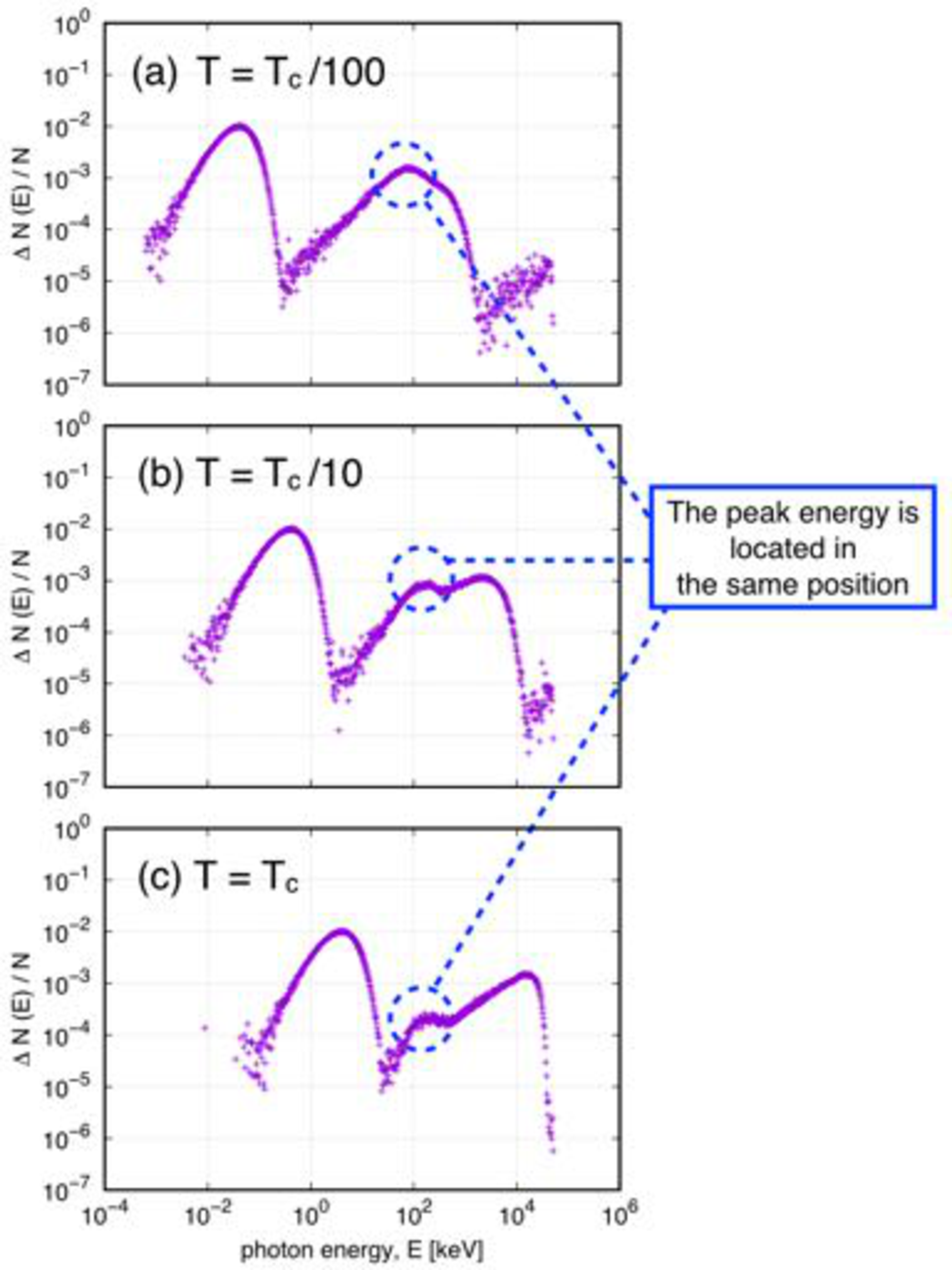}
    \caption{Spectra with different temperatures. (a) Spectrum with hundredth part of the criterial temperature, $T_c$, (b) spectrum with tenth part of $T_c$, and (c) spectrum with $T_c$.}
    \label{fig:spec_T-dif}
  \end{center}
\end{figure}

In order to show that the peak energy is located at $\sim$$10^2\ {\rm keV}$ under any condition, 
the spectra with different temperatures are shown in Fig.~\ref{fig:spec_T-dif}.
Here, $T_c$ denotes criterial temperature. 
The position of the first peak energy in the left hand is shifted to the higher energy side by 
one order of magnitude 
as the temperature increases by 
one order of magnitude 
(from (a) to (c)).
However, the peak of $\sim$$10^2\ {\rm keV}$ remains 
since the peak position is determined by the photon energy corresponding to the electron rest mass, 
and the photon energy is unchanged as long as being calculated in the same inertial frame.

\section{Post-processed radiative transfer computation of relativistic hydrodynamics simulation}
\label{sec:couple-hydro}

It is challenging task to reproduce the shock wave with $\Gamma_s = 100$ in numerical simulation 
with any numerical schemes. 
So, for developing the relativistic radiation hydrodynamics code, it is necessary to develop not only the MC radiative transfer code 
but also hydrodynamics simulation code which is appropriate for high $\Gamma_s$; 
however, it is out of the scope for this article, 
and we show the result of radiative transfer computation on the relativistic hydrodynamical flow-field 
with $\Gamma_s = 100$ only as a guide.
\begin{figure}[t]
  \begin{center}
    \vspace{-10mm}
    \includegraphics[width=0.8\linewidth]{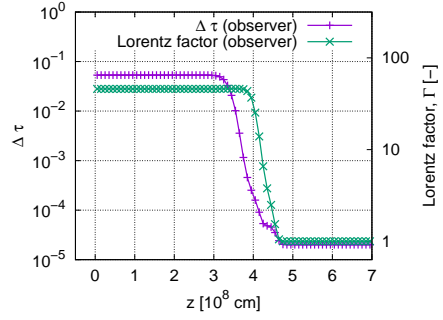}
    \vspace{10 mm}
    \caption{The distributions of optical depth and Lorentz factor of the flow velocity in the hydrodynamical flow-field at a certain time of $t = 0.01$ s.}
     \label{fig:hydro_dens_velo}
  \end{center}
\end{figure}

We performed post-processed computation of the MC radiative transfer 
on the time-dependent flow-field with relativistic hydrodynamics computation 
but no feedback from interaction of photon with flow-field matter.
Relativistic hydrodynamics simulation code was the same as that in the previous work \cite{nagakura}.
Hydrodynamics simulation was performed in advance, 
and radiative transfer was computed as a post-process.
The computation was conducted on a one-dimensional system 
with $10^5$ uniform computational cells in the $z$-direction 
using $10^5$ sample particles 
for comparison with the model computations presented in Sec.~\ref{sec:smeared_discontinuous}.
The initial conditions were the same as shown in Table~\ref{tab:butsuriryou} 
obtained by analytically solving R-H relations, 
and the velocities were transformed from the shock rest frame to the shock moving frame with $\Gamma_s = 100$.
The shock front was initially located near the left boundary.
The computation was terminated when the shock front reaches at the right boundary.
The distributions of optical depth and Lorentz factor of the flow velocity in the hydrodynamical flow-field 
at a certain time of $t = 0.01$ s 
are shown in Fig.~\ref{fig:hydro_dens_velo}. 
The smeared shock front is similar to that in the modeled flow-field as shown in Fig.~\ref{fig:ini_dens_velo_d-0-4}~(b).

\begin{figure}[t]
  \begin{center}
    \vspace{-10mm}
    \includegraphics[width=0.8\linewidth]{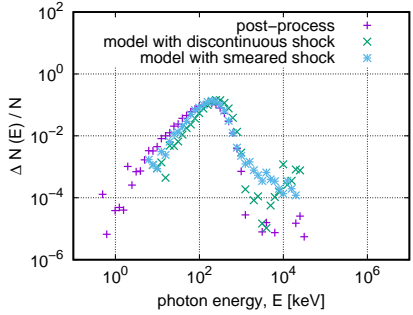}
    \vspace{10 mm}
    \caption{Comparison of spectra obtained from post-processed radiative transfer computation 
        in time-dependent background flow-field by relativistic hydrodynamics simulation 
        with spectra obtained in the model computation. 
        The models include both discontinuous and smeared shock fronts.}
     \label{fig:spec_couple}
  \end{center}
\end{figure}

Fig.~\ref{fig:spec_couple} shows the photon spectrum produced for two types of models.
Models labeled with ``model with discontinuous shock'' and ``model with smeared shock'' refer to 
calculations in which the modeled background flow-field is set up as in Sec.~\ref{sec:smeared_discontinuous}. 
The time interval for updating the flow-field in the model computations 
are set as satisfying $\alpha \sim 10^{-1}$ 
in the situation with $\Gamma_a = 100$ in Sec. \ref{sec:const_dt}; 
that is, $\Delta t \sim 10^{-4}$~s. 
%
%
The other spectrum of Fig.~\ref{fig:spec_couple} 
corresponds to post-processed radiative transfer computation of relativistic hydrodynamics simulation 
in which the background flow-field is evolved employing the hydrodynamical time step $\Delta t$; 
namely, $\Delta t \sim 10^{-3}$ s.
We note that the value of the hydrodynamical time step is larger by one order of magnitude 
than in the model computations shown in Sec.~\ref{sec:smeared_discontinuous}.
The reason to set such large value of $\Delta t$ is to limit the computational cost. 
The spectrum in the high-energy side has the positive-index power-law 
in the model computation with the discontinuous shock front, 
while the negative-index power-law appears in the model computation with the smeared shock front.
The negative power-law slope becomes steeper as the shock width increases 
because bulk Compton scattering rarely occurs as mentioned in Sec.~\ref{sec:shock-width}.
In the hydrodynamics computation, since the shock width becomes wider due to numerical diffusion as time advances, 
the high-energy photons further decrease.
The spectrum obtained from the post-processed simulation 
has the small number of photons in the high-energy side, 
and this feature is consistent with the model computations. 

In the numerical simulation, the flow-field cannot keep a single shock structure 
even when the initial conditions obtained by analytically solving R--H relations are employed.
The R--H equations are the relation that links between the upstream and downstream sides of the shock wave. 
So, if the flow-field structure is calculated analytically with the same initial conditions, 
it can keep a single shock structure. 
This difference between analytical solution and numerical one is caused 
by using the approximate Riemann solver in the numerical simulation. 
Therefore, the flow-field evolves to the Riemann problem as shown in Fig.~\ref{fig:hydro_dens_velo}.
The realistic flow-field developing with time cannot be represented by the simple model flow-field. 
So, the shock structure in our model flow-field corresponds to a snapshot of the flow-field 
computed by hydrodynamics computation at an early time. 
For the above reason, there is the lag between two jumps of the optical depth and the velocity as shown in Fig.~\ref{fig:hydro_dens_velo}. 
The optical depth goes up in the region which already reaches the terminal (downstream) velocity where the velocity does not significantly change. 
So, a large number of photons cannot experience the difference in the bulk Lorentz factor in the two scattering positions. 
This should kill lots of the high energy photons produced by the bulk Compton scattering which can be produced in the model computation 
with the discontinuous shock front.

Therefore, radiation hydrodynamics computation with coarse computational grids 
resulting in the expanded shock width 
and with the lag between the optical depth and the velocity jumps
cannot produce a lot of high-energy photons 
due to less bulk Compton scattering. 
The effect of the simulation condition on the spectrum is sustained 
in the result of the test calculation performed in this paper.



\vspace{20mm}
\begin{thebibliography}{00}


\bibitem{pomraning}
  G. C. Pomraning, The Equations of Radiation Hydrodynamics, Pergamon Press, California (1973).
\bibitem{webb}
  G. M. Webb, Astrophys. J. 296 (1985) 319.
\bibitem{kirk}
  J. G. Kirk, D. B. Melrose, and E. R. Priest, Plasma Astrophysics, Springer-Verlag, Berlin (1994).
\bibitem{anderson}
  J. L. Anderson and E. A. Spiegel, Astrophys. J. 171 (1972) 127.

\bibitem{tominaga}
  N. Tominaga, S. Shibata, and S. I. Blinnikov, Astrophys. J. Suppl. Ser. 219 (2015) 38.

\bibitem{rees}
  M. J. Rees and P. Meszaros, Astrophys. Lett. 430 (1994) L93.

\bibitem{sari}
  R. Sari and T. Piran, Astrophys. J. 485 (1997) 270.
\bibitem{kobayashi}
  S. Kobayashi, T. Piran, and R. Sari, Astrophys. J. 490 (1997) 92.
\bibitem{mimica1}
  P. Mimica and M. A. Aloy, Mon. Not. R. Astron. Soc. 401 (2010) 525.
\bibitem{mimica2}
  P. Mimica and M. A. Aloy, Mon. Not. R. Astron. Soc. 421 (2012) 2635.
\bibitem{eichler}
  D. Eichler and A. Levinson, Astrophys. J. 529 (2000) 146.

\bibitem{meszaros}
  P. Meszaros and M. J. Rees, Astrophys. J. 530 (2000) 292.

\bibitem{aloy1}
  M. A. Aloy et al., Astrophys. J. Lett. 531 (2000) L119. 
\bibitem{aloy2}
  M. A. Aloy et al., Astron. Astrophys. 396 (2002) 693.
\bibitem{zhang}
  W. Zhang, S. E. Woosley, and A. I. Macfadyen, Astrophys. J. 586 (2003) 356.

\bibitem{mizuta1}
  A. Mizuta et al., Astrophys. J. 651 (2006) 960.

\bibitem{nagakura}
  H. Nagakura et al., Astrophys. J. 731 (2011) 80.

\bibitem{matsumoto}
  J. Matsumoto and Y. Masada, Astrophys. J. Lett. 772 (2013) L1.

\bibitem{lazzati}
  D. Lazzati, B. J. Morsony, and M. C. Begelman, Astrophys. J. 700 (2009) L47.

\bibitem{mizuta2}
  A. Mizuta, S. Nagataki, and J. Aoi, Astrophys. J. 732 (2011) 26.
\bibitem{martinez1}
  C. Cuesta-Martinez, M. A. Aloy, and P. Mimica, Mon. Not. R. Astron. Soc. 446 (2015) 1716.
\bibitem{martinez2}
  C. Cuesta-Martinez et al., Mon. Not. R. Astron. Soc. 446 (2015) 1737. 
\bibitem{briggs}
  M. S. Briggs et al., Astrophys. J. 524 (1999) 82.
\bibitem{janka0}
  H. T. Janka, Nuclear Astrophysics: Proceedings of the Workshop 287 (2005) 319.
\bibitem{janka1}
  H. T. Janka and W. Hillebrandt, Astron. Astrophys. Suppl. Ser. 78 (1989) 375.
\bibitem{janka2}
  H. Th. Janka, Astron. Astrophys. 256 (1992) 452.
\bibitem{keil}
  M. Th. Keil and G. G. Raffelt, Astrophys. J. 590 (2003) 971.
\bibitem{abdikamalov}
  E. Abdikamalov et al., Astrophys. J. 755 (2012) 111.
\bibitem{maeda}
  K. Maeda, Astrophys. J. 644 (2006) 385.

\bibitem{suzuki}
  A. Suzuki and T. Shigeyama, Astrophys. J. 719 (2010) 881.

\bibitem{lucy}
  L. B. Lucy, Astronomy \& Astrophys. 429 (2005) 19.

\bibitem{beloborodov}
  A. M. Beloborodov, Astrophys. J. 737 (2011) 68.

\bibitem{ayako1}
  A. Ishii et al., High Energ. Dens. Phys. 9 (2013) 280.

\bibitem{shibata}
  S. Shibata, N. Tominaga, and M. Tanaka, Astrophys. J. Lett. 787 (2014) L4. 

\bibitem{ayako2}
  A. Ishii et al., High Energ. Dens. Phys., (2014).

\bibitem{ryde}
  F. Ryde and A. Pe'er, Astrophys. J. 702 (2009) 1211.
\bibitem{thone}
  C. C. Th\"{o}ne et al., Nature 480 (2011) 72.
\bibitem{peer}
  A. Pe'er and F. Ryde, Astrophys. J. 732 (2011) 49. 
\bibitem{giannios1}
  D. Giannios and H. C. Spruit, Astron. Astrophys. 469 (2007) 1.
\bibitem{giannios2}
  D. Giannios, Astron. Astrophys. 480 (2008) 305.
\bibitem{nakayama}
  K. Nakayama and T. Shigeyama, Astrophys. J. 627 (2005) 310.

\bibitem{takagi}
  R. Takagi and S. Kobayashi, Astrophys. J. 622 (2005) L25.

\bibitem{ohtani}
  Y. Ohtani, A. Suzuki, and T. Shigeyama, Astrophys. J. 777 (2013) 113. 

\bibitem{ito}
  H. Ito et al., Astrophys. J. 777 (2013) 62.

\bibitem{ito2}
  H. Ito et al., Astrophys. J. 789 (2014) 159. 

\bibitem{noebauer}
  U. M. Noebauer et al., Mon. Not. R. Astron. Soc. 425 (2012) 1430.

\bibitem{ito3}
  H. Ito et al., Astrophys. J. Lett. 814 (2015) L29. 

\bibitem{roth}
  N. Roth and D. Kasen, Astrophys. J. Suppl. Ser. 217 (2015) 9.
\bibitem{mimica3}
  P. Mimica et al., Astron. Astrophys. 441 (2005) 103.
\bibitem{fleck}
  J. A. Fleck JR. and J. D. Cummings, J. Comput. Phys. 8 (1971) 313.
\bibitem{mcclarren}
  R. G. McClarren and T. J. Urbatsch, J. Comput. Phys. 228 (2009) 5669.
\bibitem{gentile}
  N. A. Gentile, J. Comput. Phys. 172 (2001) 543.
\bibitem{wollaeger}
  R. T. Wollaeger et al., Astrophys. J. Suppl. Ser. 209 (2013) 36.
\bibitem{cleveland}
  M. A. Cleveland and N. Gentile, J. Comput. Phys. 291 (2015) 1.
\bibitem{densmore1}
  J. D. Densmore and E. W. Larsen, J. Comput. Phys. 291 (2015) 1.
\bibitem{densmore2}
  J. D. Densmore et al., J. Comput. Phys. 284 (2015) 40.
\bibitem{heitler}
  W. Heitler, The Quantum Theory of Radiation, Dover Publications, New York (2010).
\bibitem{schutz}
  B. Schutz, A First Course in General Relativity, Cambridge University Press, New York (2009).

\bibitem{rybicki}
  G. B. Rybicki and A. P. Lightman, Radiative Processes in Astrophysics, Wiley-VCH, M\"{o}rlenbach (1985).

\end{thebibliography}
\end{document}